\documentclass[pra,aps,10pt,superscriptaddress,twocolumn,floatfix]{revtex4-1}
\usepackage{graphicx,color}
\usepackage[nice]{nicefrac}							

\usepackage{diagbox}
\usepackage{amsmath,amssymb}
\usepackage{tikz}
\usetikzlibrary{decorations.markings}
\usetikzlibrary{arrows}
\usetikzlibrary{decorations.pathreplacing}
\usetikzlibrary{calc,intersections,through,backgrounds}
\usepackage[bb=boondox]{mathalfa}
\usepackage{braket}		
\usepackage{mathtools}
\usepackage{makecell}
\usepackage{multirow}
\usepackage{layouts}
\usepackage{tikz}
\usetikzlibrary{decorations.pathmorphing}
\usetikzlibrary{arrows}
\usetikzlibrary{decorations.markings}

\tikzset{db/.style={double distance = 1pt,line cap=rect}}
\tikzset{->-/.style={
         postaction={decoration={markings,mark=at position 0.54 with {\arrow[line width=0.1mm,scale=1.5]{>}}},decorate}
	}}
\tikzset{-<-/.style={
         postaction={decoration={markings,mark=at position 0.5 with {\arrow[line width=0.1mm,scale=1.5]{<}}},decorate}
	}}
\tikzset{<->/.style={
         postaction={decoration={markings,mark=at position 0.25 with {\arrow[line width=0.1mm,scale=1.5]{<}},mark=at position 0.75 with {\arrow[line width=0.1mm,scale=1.5]{>}}},decorate}
	}}

\newcommand{\ccaption}[2]{\caption[#1]{\textit{#1.} #2}}
\newcommand{\ii}{\mathrm{i}} 
\newcommand{\ie}{{\it i.e.},\ }
\newcommand{\eg}{{\it e.g.},\ }

\newcommand{\id}{\mathbb{1}}

\makeatletter
\renewcommand*\env@matrix[1][\arraystretch]{%
  \edef\arraystretch{#1}%
  \hskip -\arraycolsep
  \let\@ifnextchar\new@ifnextchar
  \array{*\c@MaxMatrixCols c}}
\makeatother

\usepackage[plainpages=false,pdfpagelabels,colorlinks=true,linkcolor=red,urlcolor=blue,citecolor=blue,pdftitle={},pdfauthor={},pdfdisplaydoctitle=true,pdfduplex=DuplexFlipLongEdge]{hyperref}

\definecolor{darkred}{rgb}{0.90,0.2,0.2}
\definecolor{darkgreen}{rgb}{0,0.60,.2}
\definecolor{darkblue}{rgb}{0.1,0.3,1}
\definecolor{grey}{cmyk}{0,0,0,0.25}
\definecolor{orange}{cmyk}{0,0.6,0.8,0}

\newcommand{\la}[1]{{\color{OliveGreen}#1}}

\definecolor{OliveGreen}{HTML}{009900}
\definecolor{lred}{rgb}{0.6,0,0}
\definecolor{oorange}{HTML}{e55c0e}
\definecolor{lgrey}{gray}{0.9}
\definecolor{brown}{HTML}{84554d}
\definecolor{dred}{rgb}{0.9,0,0}
\definecolor{dblue}{rgb}{0,0,0.6}
\definecolor{green}{rgb}{0,0.6,0}
\definecolor{dgreen}{rgb}{0,0.4,0}
\definecolor{pink}{RGB}{214,18,255}
\definecolor{G}{rgb}{0.6,0,0}
\definecolor{J}{rgb}{0.6,0,0.6}
\definecolor{Om}{rgb}{0,0,0.6}
\definecolor{cit}{rgb}{0,0.7,0}

\newcommand{\be}[1]{{\color{lred}#1}}

\usetikzlibrary{calc,intersections,through,backgrounds}
\newcommand\manifold[3][]{
	\draw[every to/.style={out=-20,in=160,relative},#1] (#2) 
	to ($(#2 -| #3)!0.2!(#2 |- #3)$)
	to (#3)
	to ($(#2 -| #3)!0.8!(#2 |- #3)$)
	to cycle;
}

\begin{document}

\title{Gaussian time dependent variational principle for the Bose-Hubbard model}

\author{Tommaso Guaita}
\email{tommaso.guaita@mpq.mpg.de}
\affiliation{Max  Planck  Institute of Quantum Optics, Hans-Kopfermann-Stra{\ss}e 1, D-85748 Garching bei M\"unchen, Germany}
\affiliation{Munich Center for Quantum Science and Technology, Schellingstra{\ss}e 4, D-80799 M\"{u}nchen, Germany}
\author{Lucas Hackl}
\email{lucas.hackl@mpq.mpg.de}
\affiliation{Max Planck  Institute of Quantum Optics, Hans-Kopfermann-Stra{\ss}e 1, D-85748 Garching bei M\"unchen, Germany}
\affiliation{Munich Center for Quantum Science and Technology, Schellingstra{\ss}e 4, D-80799 M\"{u}nchen, Germany}
\author{Tao Shi}
\email{tshi@itp.ac.cn}
\affiliation{CAS Key Laboratory of Theoretical Physics, Institute of Theoretical Physics, Chinese Academy of Sciences, Beijing 100190, China}
\affiliation{CAS Center for Excellence in Topological Quantum Computation, University of Chinese Academy of Sciences, Beijing 100049, China}
\author{Claudius Hubig}
\affiliation{Max  Planck  Institute of Quantum Optics, Hans-Kopfermann-Stra{\ss}e 1, D-85748 Garching bei M\"unchen, Germany}
\affiliation{Munich Center for Quantum Science and Technology, Schellingstra{\ss}e 4, D-80799 M\"{u}nchen, Germany}
\author{Eugene Demler}
\affiliation{Lyman Laboratory, Department of Physics, Harvard University, 17 Oxford St., Cambridge, MA 02138, USA}
\author{J. Ignacio Cirac}
\affiliation{Max  Planck  Institute of Quantum Optics, Hans-Kopfermann-Stra{\ss}e 1, D-85748 Garching bei M\"unchen, Germany}
\affiliation{Munich Center for Quantum Science and Technology, Schellingstra{\ss}e 4, D-80799 M\"{u}nchen, Germany}

\begin{abstract}
We systematically extend Bogoliubov theory beyond the mean field approximation of the Bose-Hubbard model in the superfluid phase. Our approach is based on the time dependent variational principle applied to the family of all Gaussian states (\ie Gaussian TDVP). First, we find the best ground state approximation within our variational class using imaginary time evolution in 1d, 2d and 3d. We benchmark our results by comparing to Bogoliubov theory and DMRG in 1d. Second, we compute the approximate 1- and 2-particle excitation spectrum as eigenvalues of the linearized projected equations of motion (linearized TDVP). We find the gapless Goldstone mode, a continuum of 2-particle excitations and a doublon mode. We discuss the relation of the gap between Goldstone mode and 2-particle continuum to the excitation energy of the Higgs mode. Third, we compute linear response functions for perturbations describing density variation and lattice modulation and discuss their relations to experiment. Our methods can be applied to any perturbations that are linear or quadratic in creation/annihilation operators. Finally, we provide a comprehensive overview how our results are related to well-known methods, such as traditional Bogoliubov theory and random phase approximation.
\end{abstract}

\maketitle

\section{Introduction}
The Bose-Hubbard model provides a theoretical description of interacting cold atoms in optical lattices~\cite{bloch_ultracold_gases_2008}, which in the last years have proven to be a promising experimental platform. Its Hamiltonian is given by
\begin{align}
    \hat{H}&=-\sum_{\left\langle i,j\right\rangle} \hat{b}_i^\dag \hat{b}_j + \frac{U}{2}\sum_i \hat{b}_i^\dag \hat{b}_i^\dag \hat{b}_i \hat{b}_i - \mu \sum_i \hat{b}_i^\dag \hat{b}_i\,,\label{eq:BH_Hamiltonian}
\end{align}
where $\hat{b}^\dag_i$ and $\hat{b}_i$ are the bosonic creation and annihilation operators for a particle on site $i$ of a square lattice. The model has been analyzed theoretically with a several different methods, ranging from the historical Bogoliubov theory~\cite{Bogolyubov_1947} to later approaches based on the Gutzwiller ansatz~\cite{pekker_signatures_2012}.

For different choices of the model parameters $U$ and $\mu$, the system exhibits two different phases in the thermodynamic limit: a superfluid phase (small $U$) and a Mott insulator phase (large $U$). One characterization of the superfluid phase is that the $\mathrm{U}(1)$ symmetry generated by the particle number operator $\hat{N}=\sum_i \hat{b}_i^\dag \hat{b}_i$~\cite{sachdev_quantum_1999} is spontaneously broken for $N\to\infty$. This leads to both a gapless Goldstone mode and a massive Higgs amplitude mode in the excitation spectrum around the transition. The properties of these have both been described theoretically, \eg with methods based on the Gutzwiller ansatz~\cite{huber_amplitude_mode_2008, huber_dynamical_properties_2007, bissbort_quasi-particle_2014}, strong coupling~\cite{sengupta_mott-insulatorsuperfluid_2005,fitzpatrick_contour-time_2018}, the variational cluster approach~\cite{knap_variational_2011}, the random phase approximation~\cite{menotti_spectral_weight_2008} or the ladder diagram approximation for the continuum theory~\cite{stoof_variational_1997}, and observed in experimental realizations of the model~\cite{endres2012higgs,bissbort_detecting_2011,shori_excitations_2004}.

The aim of our paper is to introduce a systematic generalization of the Bogoliubov mean field theory for the superfluid phase. Our method is best described as Gaussian time dependent variational principle (Gaussian TDVP), \ie we compute system properties from the family of bosonic Gaussian states~\cite{weedbrook_gaussian_2012} given by displaced and squeezed vacua. This is in contrast to Bogoliubov theory which is based on the smaller variational family of coherent states, \ie just displaced vacua.

Bogoliubov theory describes the model by suitably truncating the Hamiltonian to a quadratic non-interacting mean field Hamiltonian. The minimal energy of this Hamiltonian approximates remarkably well the exact ground state energy. Furthermore, the mean field Hamiltonian can be diagonalized using Bogoliubov transformations and its spectrum describes the dispersion relation of the gapless Goldstone mode of the model. This last step is equivalent to applying the linearized time dependent variational principle to coherent states (coherent TDVP).

Bogoliubov theory, however, also presents several drawbacks. First, the Bogoliubov ground state energy approximation is not variational, \ie the mean field ground state does not minimize the expectation value with respect to the full Hamiltonian. Second, it does not capture other excitations beyond the Goldstone one, such as the Higgs amplitude mode or bound doublon states. Third, the Goldstone quasiparticles are non-interacting and thus, the decay of quasiparticles excitations can only be studied by re-including the initially discarded Hamiltonian terms as a perturbation~\cite{beliaev-energy-1958}.

By instead applying linearized TDVP to an extended variational manifold, \ie the larger class of Gaussian states in place of just coherent states, we overcome all of these drawbacks. First, we compute a variational ground state approximation given by the Gaussian state $\ket{\psi_{\mathrm{g}}}$ with minimal energy expectation value. For this, we use imaginary time evolution and show that $\ket{\psi_{\mathrm{g}}}$ can be efficiently computed in any dimension from two self-consistent equations. Second, our approximate excitation spectrum captures both 1- and 2-particle states, which include the gapless Goldstone mode, a doublon mode and a gapped mode which may be interpreted as a Higgs amplitude mode. The approximate excitation spectrum arises as the eigenvalues of the linearized TDVP equations of motion on the tangent space of Gaussian states. Third, the Gaussian tangent plane naturally captures the interaction of quasi-particle excitations in the 1- and 2-particle sector. This allows us to extract spectral response functions associated to linear and quadratic perturbations and to compute decay and time evolution of excitations.\\

This paper is structured as follows: In Section~\ref{sec:Gaussian_ground_state_approximation}, we introduce our variational manifold and compute the best approximation of the system's ground state in the superfluid phase, \ie the Gaussian state with the minimal energy expectation value.
In Section~\ref{sec:quasiparticle_excitations}, we study the linearization of the projected real time evolution on such manifold to obtain an expression for the system's excitation spectrum.
In Section~\ref{sec:linear_response}, we develop a geometric linear response theory consistent with our approximation scheme to capture how linear perturbations couple to different parts of the spectrum.
In Section~\ref{sec:comparison}, we  expand on the relationship between our methods and others also based on a Gaussian or coherent state ansatz manifold (specifically Bogoliubov theory) and discuss differences and advantages.
Finally, we conclude in Section~\ref{sec:discussion} with a comprehensive discussion of our results. In Appendices~\ref{app:Bogoliubov}-\ref{app:iterated_Bogoliubov}, we review Bogoliubov theory, its partial equivalence to coherent TDVP and how to make it self-consistent by iteration. In Appendices~\ref{app:approximate_ground_state}-\ref{app:linear_response_theory}, we provide further details on the Gaussian ground state approximation, the linearized equations of motion and linear response theory. Finally, in Appendix~\ref{app:feynman_diagrams} we illustrate the equivalence between our Gaussian method and the random phase approximation scheme based on ladder Feynman diagrams.

\section{Gaussian ground state approximation}\label{sec:Gaussian_ground_state_approximation}
As first step of applying our variational methods, we compute the best Gaussian state $\ket{\psi_{\mathrm{g}}}$, \ie the normalized Gaussian states whose energy expectation value $E_{\ket{\psi_{\mathrm{g}}}}=\bra{\psi_{\mathrm{g}}}\hat{H}\ket{\psi_{\mathrm{g}}}$ on the full Hamiltonian is minimal.

\subsection{Variational manifold}
We generalize the Bogoliubov theory of the Bose-Hubbard model by extending the variational manifold for the system state to the full manifold $\mathcal{M}$ of bosonic Gaussian states. This is in contrast to regular Bogoliubov theory, where the variation is only done with respect to coherent states. The manifold of Gaussian states can be conveniently parametrized by first squeezing and then displacing the reference vacuum $\ket{0}$, \ie we consider the variational manifold
\begin{align}
    \mathcal{M}=\Big\{\ket{\beta,\lambda}=\mathcal{U}(\beta,\lambda)\ket{0}\Big\}\,,
    \label{eq:gaussian_manifold}
\end{align}
with unitaries $\mathcal{U}(\beta,\lambda)=\mathcal{D}(\beta)\mathcal{S}(\lambda)$ defined by
\begin{align}
        \mathcal{D}(\beta)&= \exp\left[\frac{1}{2}\sum_k\left(\beta_k \hat{b}^\dag_k - \beta^*_k \hat{b}_k\right)\right]\,,\\
        \mathcal{S}(\lambda)&=\exp\left[\frac{1}{2}\sum_{kq} \left(\lambda_{k,q} \hat{b}^\dag_{k-q} \hat{b}^\dag_q - \lambda^*_{k,q} \hat{b}_{k-q} \hat{b}_q\right)\right]\,,
        \label{eq:gaussian states}
\end{align}
where $\hat{b}_k=\frac{1}{\sqrt{N}} \sum_i e^{-\ii k x_i} \hat{b}_i$ are the momentum space annihilation operators. Here, $\beta_k$ is a complex vector and $\lambda_{k,q}$ is a complex matrix invariant under $q\to k-q$. The indices $k$ and $q$ run in the reciprocal lattice. The only redundancy contained in this parametrization is the symmetry of $\lambda$. For a system with $N$ bosonic degrees of freedom, we count $N(N+3)/2$ complex coordinates $(\beta_k,\lambda_{k,q})$ or $N(N+3)$ real coordinates
\begin{align}
    x^a=\big(\mathrm{Re}(\beta_k),\mathrm{Re}(\lambda_{k,q}),\mathrm{Im}(\beta_k),\mathrm{Im}(\lambda_{k,q})\big)\,.\label{eq:real_coordinates}
\end{align}
We will use the shorthand notation $\mathcal{U}(x^{\mathrm{g}})$ for the choices of $x^a$, such that $\mathcal{U}(x^{\mathrm{g}})\ket{0}=\ket{\psi_{\mathrm{g}}}$.

The manifold is closed under the action of any subgroup generated by any operators that are linear and quadratic in creation/annihilation operators. In particular, this applies to the $\mathrm{U}(1)$ symmetry group generated by the total number operator $\hat{N}=\sum_i\hat{b}^\dag\hat{b}_i$. Here, for any Gaussian state $\ket{\psi}$ other than the vacuum $\ket{0}$ we find a whole ring of inequivalent states $e^{\ii \theta \hat{N}}\ket{\psi}$ with the same energy expectation value. Therefore, we expect this variational manifold to be well suited to capture spontaneously broken $\mathrm{U}(1)$ symmetry phase of the system, \ie the superfluid phase, and its features, such as the massless Goldstone mode. 

While the symmetry of the Bose-Hubbard model is known to be only spontaneously broken in the thermodynamic limit ($N\to\infty$), our ansatz already gives rise to a family of non-symmetric approximate ground states at finite $N$. We furthermore point out that, while finite temperature spontaneous breaking of a continuous symmetry at zero temperature is ruled out in 1D, there might still be quasi-long range order, therefore a broken symmetry ansatz can turn out to be a reasonable choice also in 1D.

The manifold contains a submanifold of states which are translationally invariant, namely the set of states $\ket{\psi(\beta_k,\lambda_{k,q})}$ with $\beta_k=\delta_{k,0} \beta_0$ and $\lambda_{k,q}=\delta_{k,0} \lambda_{0,q}$.
For the ground state search, it is sufficient to restrict ourselves to this submanifold as we expect the ground state to preserve the translational symmetry of the problem.
For the study of excitations around the translationally invariant ground state, we will then use the full manifold in order to capture also excitations with non-zero momentum.

The tangent space $\mathcal{T}_{\ket{\psi}}$ of the variational manifold at the point $\ket{\psi}$ is naturally spanned by the states with 1- and 2-particle excitations, \ie
\begin{align}
     \mathcal{T}_{\ket{\psi(x)}}\mathcal{M}=\mbox{span} {\left\{ \:\mathcal{U}(x)\hat{b}^\dag_k\ket{0},\;\mathcal{U}(x)\hat{b}^\dag_{k-q} \hat{b}^\dag_q\ket{0} \right\} }_{k,q}. 
     \label{eq:tangent_plane_basis}
\end{align}
Put differently, the variational class of \emph{all} Gaussian states captures accurately the 1- and 2-particle quasiparticle excitation sector of our model.

\subsection{Imaginary time evolution}
The first step of our procedure to exploit the given choice of variational manifold is to find within it the best approximation of the ground state, that is the state with the lowest energy expectation value. One strategy to do this is to consider the projected imaginary time evolution. This is the solution of the evolution equation
\begin{align}
    \frac{d}{d\tau}\ket{\psi(\tau)}=- \mathbb{P}_{\ket{\psi(\tau)}} \hat{H} \ket{\psi(\tau)},
\end{align}
where $\mathbb{P}_{\ket{\psi(\tau)}}$ is the orthogonal projector onto the tangent space to the manifold at $\ket{\psi(\tau)}$. This projection ensures that the solution will be contained in the ansatz manifold at all times $\tau$. This evolution converges from a random initial state to a local minimum of the energy expectation value function and can be shown to be equivalent to a gradient descent method. For Gaussian states, we find simple equations for the stationary point of this evolution, \ie the state $\ket{\psi_{\mathrm{g}}}\in\mathcal{M}$ such that $-\mathbb{P}_{\ket{\psi_{\mathrm{g}}}}\hat{H} \ket{\psi_{\mathrm{g}}}=0$, and see that they only admit a single solution up to the redundancy generated by $e^{\ii \alpha\hat{N}}$.

This solution $\ket{\psi_{\mathrm{g}}}$ can be characterized analytically, independently of the system size or dimensionality, in terms of two parameters $A$ and $B$, which can be efficiently computed numerically as the fixed point of two coupled self-consistent equations. For more details on this calculation and on how to parametrize the approximate ground state see Appendix \ref{app:approximate_ground_state}.

\subsection{Ground state properties}
\begin{figure}[t]
\centering
  \begin{tikzpicture}
  \draw (0,0) node[inner sep=0pt]{\includegraphics[width=\linewidth]{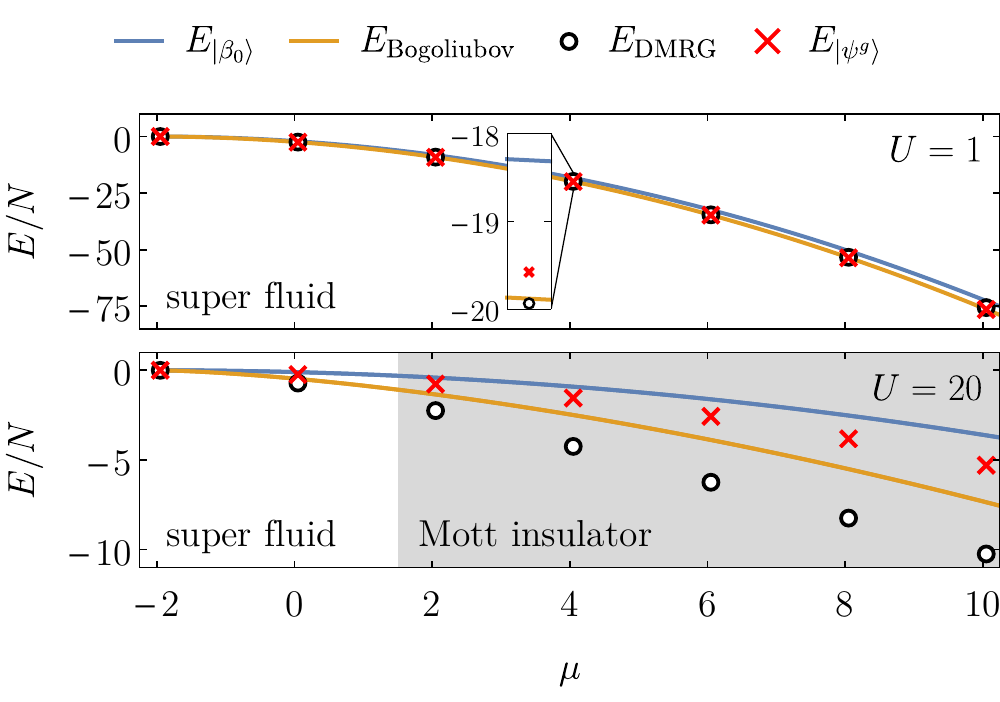}};
  \end{tikzpicture}
  \ccaption{Comparison of ground state energies in 1d}{We compare the following approaches: (a) Minimal energy on manifold of coherent states $E_{\ket{\beta^c_0}}$ from~\eqref{eq:E_beta0}, (b) Bogoliubov ground state energy $E_{\mathrm{Bogoliubov}}=E_{\ket{\beta^c_0}}-\Delta$ from~\eqref{eq:EBogo} in the limit $N\to\infty$, (c) DMRG energy $E_{\mathrm{DMRG}}$ and (d) minimal energy $E_{\ket{\psi^g}}$ of all Gaussian states for $N=501$. The DMRG results were computed for finite systems with open boundary conditions and then extrapolated to the thermodynamic limit $N\to\infty$. The Gaussian state energy $E_{\ket{\psi^g}}$ at $N=501$ appears to have already substantially reached the thermodynamic limit value.}
  \label{fig:comparison_Gaussian_vs_DMRG}
\end{figure}

Having obtained an analytical expression for the approximate ground state, it is then possible to calculate the predictions of our model for ground state properties such as the energy and particle densities. The quality of our method can be benchmarked by comparing these quantities with the ones obtained through other methods, such as Bogoliubov theory or, at least in one dimension, with a numerical DMRG~\cite{white1992density} calculation (see Figure \ref{fig:comparison_Gaussian_vs_DMRG}). Our variational energy $E_{\ket{\psi_{\mathrm{g}}}}=\bra{\psi_{\mathrm{g}}}\hat{H}\ket{\psi_{\mathrm{g}}}$ is higher than the DMRG one, as expected, but lower than the one obtained by other variational choices, such as the coherent state $\ket{\beta^{\mathrm{c}}_0}$ with minimal energy $E_{\ket{\beta^{\mathrm{c}}_0}}$. The energy obtained as the ground state energy of the Bogoliubov mean field Hamiltonian is generally lower than ours and remarkably close to the DMRG result. However, it is important to emphasize that this energy $E_{\mathrm{Bogoliubov}}$ is not variational as it is computed with respect to the truncated mean field Hamiltonian, which actually does not admit a well defined ground state in the zero momentum mode. More precisely the state minimizing the mean field energy is infinitely squeezed, which would lead to a diverging energy expectation value with respect to the full Bose-Hubbard Hamiltonian.

Thus, the Gaussian variational family provides a consistent class to approximate the ground state of the Bose-Hubbard model in the superfluid phase, even though its ground state energy estimate is worse than the one obtained from Bogoliubov theory. However, the strength of our extended variational family lies in its prediction of quasiparticle excitations and their properties, such as life time and linear response.

\section{Quasi particle excitations}\label{sec:quasiparticle_excitations}
We can derive an approximate excitation spectrum from the perspective of our Gaussian variational manifold by looking at real time evolution of the system projected onto the manifold. As our variational class generalizes the set of coherent state manifold used in standard Bogoliubov theory, we will be able to capture higher excitation modes of the model.

The projected real time evolution is computed as prescribed by the time dependent variational principle (TDVP)~\cite{haegeman_tdvp_2011,dirac_note_1930}. Such principle can be formulated as stating that the real time evolution projected on the manifold of Gaussian states takes the form
\begin{align}
    \frac{d}{dt}\ket{\psi(t)}= \mathbb{P}_{\ket{\psi(t)}} (-\ii \hat{H}) \ket{\psi(t)},
\end{align}
and generates a Hamiltonian time evolution flow $\Phi_t: \mathcal{M}\rightarrow\mathcal{M}$ that, in a neighbourhood of the stationary state $\ket{\psi_{\mathrm{g}}}$, reduces to a sum of phase rotations. From the perspective of our variational manifold, the frequencies of these rotations provide a natural approximation of the lowest excitation energies.

\subsection{Linearized TDVP}
\begin{figure}[t]
\centering
  \begin{tikzpicture}
  \draw (0,0) node[inner sep=0pt]{\includegraphics[width=\linewidth]{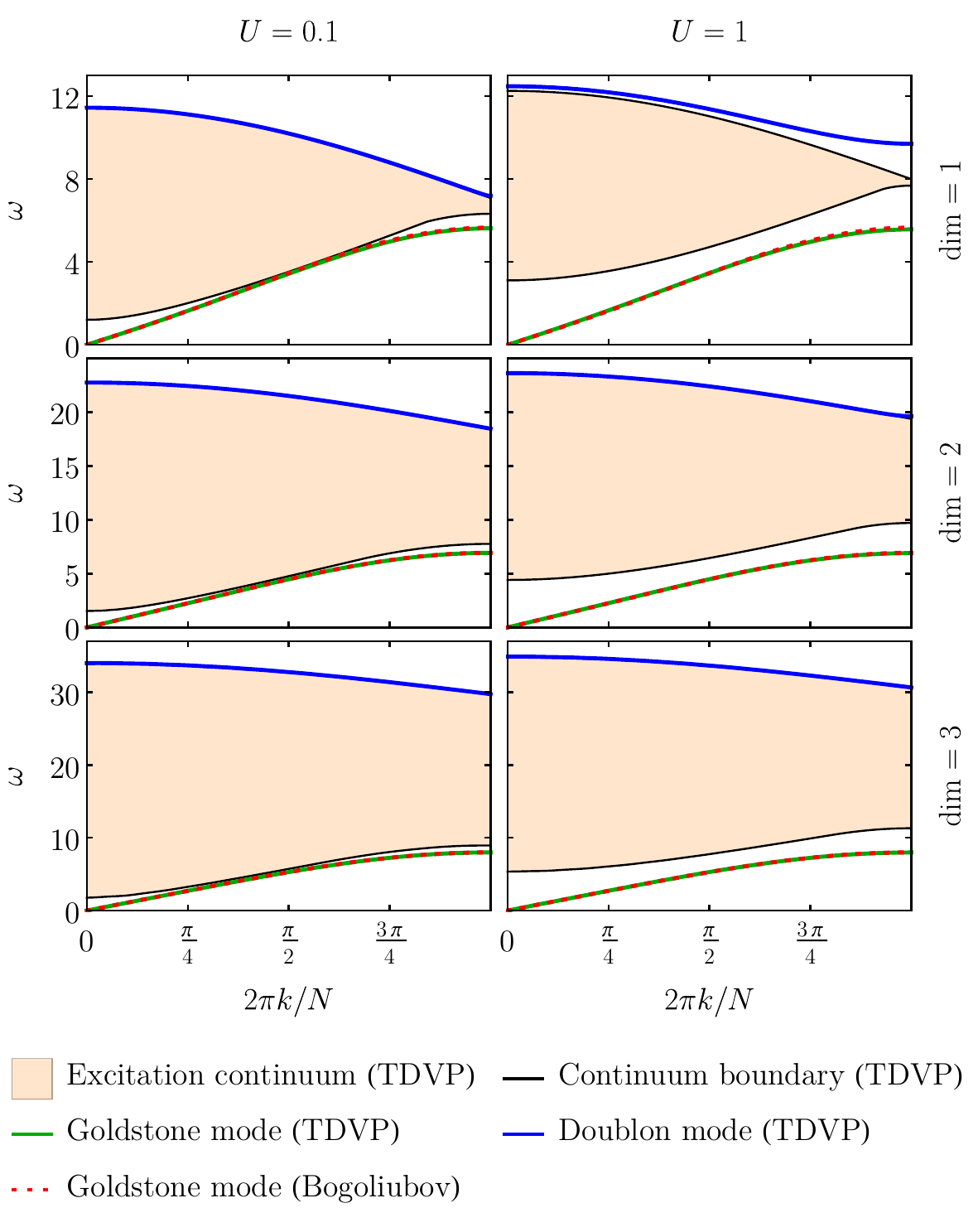}};
  \end{tikzpicture}
  \ccaption{Excitation spectra for $\mu=0$}{We compare the quasiparticle excitation spectrum computed from Gaussian TDVP with Bogoliubov theory. The results are shown for $\mu=0$ and two different values of the interaction strength ($U=0.1$ and $U=1$) in 1, 2 and 3 dimensions. The spectrum was computed as eigenvalues of $K^a{}_b$ from~\eqref{eq:Kmatrix}, where we interpolated the continuum part of the spectrum from systems The computations were performed for $N=(501,101^2,41^3)$ for $\dim=(1,2,3)$ respectively.}
  \label{fig:excitation_spectrum}
\end{figure}

We calculate the excitation energies as shown in Figure~\ref{fig:excitation_spectrum}, as the eigenvalues of the linearized equations of motion, that can be understood as a generalization of the well-known Gross-Pitaevskii equation~\cite{pitaevskii_vortex_1961, gross_structure_1961}.

In particular, we consider the linearization of this projected real time evolution around the stationary point $\ket{\psi_{\mathrm{g}}}$, that defines a linear map $K$ at the tangent space of $\ket{\psi_{\mathrm{g}}}$, mapping a tangent vector $\ket{V}$ onto
\begin{align}
    \partial_{V} \mathbb{P}_{\ket{\psi}} (-\ii \hat{H}) \ket{\psi}= \left(\partial_{V} \mathbb{P}_{\ket{\psi}}\right) (-\ii \hat{H}) \ket{\psi}+\mathbb{P}_{\ket{\psi_{\mathrm{g}}}} (-\ii \hat{H}) \ket{V}\,,\label{eq:linerized_eom}
\end{align}
where $\partial_{V}$ indicates the directional derivative on $\mathcal{M}$ evaluated at $\ket{\psi_{\mathrm{g}}}$ in the direction indicated by $\ket{V}$.

Using the fact that $\ket{\psi_{\mathrm{g}}}$ is a stationary point of the evolution, \ie $\mathbb{P}_{\ket{\psi_{\mathrm{g}}}} (-\ii \hat{H}) \ket{\psi_{\mathrm{g}}}=0$, we can write the map $K$ as a real $N(N+3)\times N(N+3)$ matrix, referring to the tangent plane basis $\{\mathcal{U}(x^{\mathrm{g}})\ket{W^a}\}$ (see Appendix~\ref{app:linearized_eom}),
\begin{align}
    K^a{}_b=\frac{\partial}{\partial x^b}\,\mathrm{Re}\bra{W^a}\mathcal{U}^\dagger(x)\hat{H}\mathcal{U}(x)\ket{0}\,,\label{eq:Kmatrix}
\end{align}
where $x^a$ refers to the real coordinates~\eqref{eq:real_coordinates} of the manifold and the orthonormal frame $\ket{W^a}$ is given by
\begin{align}
    \left\{\ket{W^a}\right\}=\left\{\hat{b}^\dag_k\ket{0},\hat{b}^\dag_{k-q}\hat{b}^\dag_{q}\ket{0},\ii\hat{b}^\dag_k\ket{0},\ii\hat{b}^\dag_{k-q}\hat{b}^\dag_{q}\ket{0}\right\}.
\end{align}
To give this matrix representation of $K$ we are here employing a real formalism in which we use a real parametrization and we consider the tangent plane as real vector space, \ie we consider the vectors $\ket{W}$ and $\ket{W'}=\ii\!\ket{W}$ to be linearly independent (and orthogonal with respect to the real inner product $\mathrm{Re}\braket{W|W'}$). This is naturally dictated by the fact that the linearized equations of motion from~\eqref{eq:linerized_eom} do not commute with complex multiplication, \ie $K^a{}_b$ is not complex-linear, as elaborated in Appendix~\ref{app:linearized_eom}. Another formal expression for the matrix $K$ is
\begin{equation}
    K^a{}_b=-\sum_c\Omega^{ac}\,\frac{\partial}{\partial x^c}\,\frac{\partial}{\partial x^b} E(x),
    \label{eq:Hessian}
\end{equation}
where $E(x)$ is the energy expectation value of the state $\ket{\psi(x)}$, and the matrix $\Omega$ is the antisymmetric symplectic form defined in~\eqref{eq:Omega}.

\begin{figure*}[t]
\centering
  \begin{tikzpicture}
  \draw (0,0) node[inner sep=0pt]{\includegraphics[width=\linewidth]{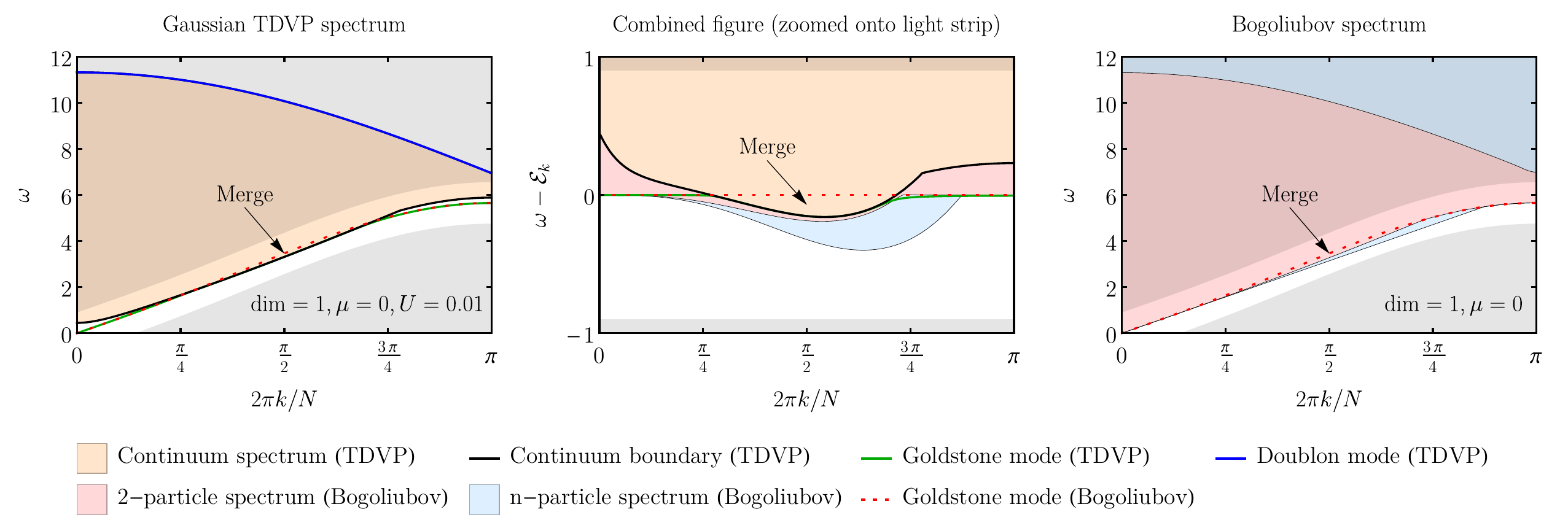}};
  \end{tikzpicture}
  \ccaption{Comparison with Bogoliubov spectrum}{We compare the spectra from Gaussian TDVP (left figure) with the one from Bogoliubov theory (right figure). For this, we overlap both figures (middle figure) and zoom into the narrow light strip around the Bogoliubov dispersion relation $\mathcal{E}_k$ (red dotted line). We see that the Goldstone mode merges into the TDVP continuum spectrum in the same region, where free Bogoliubov theory predicts $\mathcal{E}_k$ to lie inside the 2-particle continuum (indicated by arrows). The TDVP was performed in 1 dimension for $N=501$, $\mu=0$ and $U=0.01$.}  \label{fig:Bogo-comparison}
\end{figure*}

The evaluation of this matrix~\eqref{eq:Kmatrix} reduces to calculating expectation values using Wick's theorem and taking derivatives, therefore it can be calculated analytically in terms of the ground state parameters obtained in the previous section. More details on the form of $K$ can be found in Appendix \ref{app:linearized_eom}. $K$ is a symplectic matrix whose eigenvalues come in complex conjugate pairs $\pm\ii\omega$. The values of $\omega$ are our estimates of the excitation energies of the model.

Due to the translational invariance of $\hat{H}$ the matrix $K$ is block diagonal, with each block acting on the span of tangent vectors with fixed total momentum, which we labeled by $k$ in in~\eqref{eq:real_tangent_frame}. The approximate excitation energies $\omega$ can therefore also be labeled by the total momentum $k$ their respective eigenvector. The size of each block grows linearly in $N$, and therefore in the thermodynamic limit $N\to\infty$ there is an infinity of eigenvalues $\omega_k$ for each $k$, which can arrange themselves in a continuum plus possibly some discrete excitations that represent bound states.

\subsection{Excitation spectrum}
In Figure~\ref{fig:excitation_spectrum}, we show the dispersion relations obtained by diagonalizing the matrix $K$ numerically. For momenta close to zero, we always find a gapless isolated mode that agrees well with the Bogoliubov dispersion relation $\mathcal{E}_k$. However, we also find a continuum of states that have energies above this Goldstone mode and that always shows a gap around $k=0$. Finally, for certain parameter choices, \eg for strong interactions, our spectrum also contains another isolated state above the continuum, which can be interpreted as a doublon state.

We point out that the fact that our method gives a gapless mode was to be expected. Indeed, $\hat{N}$ commutes with the Hamiltonian and the vector $\hat{N}\ket{\psi}$ is part of the tangent plane for all $\ket{\psi}\in\mathcal{M}$, because $\hat{N}$ is quadratic in the bosonic creation and annihilation operators. Therefore there exists a direction in the manifold along which the energy is constant. In this direction, the Hessian $\frac{\partial}{\partial x^a}\,\frac{\partial}{\partial x^b} E(x)$ has a vanishing eigenvalue and thus, because of equation~\eqref{eq:Hessian}, also $K$ does.

Our method captures the tangent plane generated by displacements and squeezing, \ie it is spanned by 1- and 2-particle excitations. A generic eigenvector $\ket{E_k}$ of $K^a{}_b$ with momentum $k$ is
\begin{align}
    \ket{E_k}=\mathcal{U}(x^{\mathrm{g}})\,\Big[C\,\hat{b}_k^\dagger+\sum_qC_q\hat{b}^\dagger_{k+q}\hat{b}^\dagger_{-q}\Big]\ket{0}\,,
\end{align}
where $C,C_q\in\mathbb{C}$. We should therefore compare our results with the 1- and 2-particle excitation spectrum obtained from Bogoliubov theory.

Traditional Bogoliubov theory constructs the excitation spectrum from the 1-particle dispersion relation $\mathcal{E}_k$ (see~\eqref{eq:BogoEk}) of the mean field Hamiltonian
\begin{align}
    [\hat{H}]_{\ket{\beta^{\mathrm{c}}_0}}=E_{\ket{\beta^{\mathrm{c}}_0}}-\Delta^{\mathrm{c}}+\sum_{k}\mathcal{E}^{\mathrm{c}}_k\,(\delta\hat{B}^{\mathrm{c}}_k)^\dagger\delta\hat{B}_k^{\mathrm{c}}\,,
\end{align}
as reviewed in~\ref{app:Bogoliubov}. The dispersion relation $\mathcal{E}_k$ is independent of the interaction strength $U$ and becomes exact in the limit $U\to 0^+$. General eigenstates of $[\hat{H}]_{\ket{\beta^{\mathrm{c}}_0}}$ consist of non-interacting excitations created by $(\delta\hat{B}^{\mathrm{c}}_k)^\dagger$. A general 2-particle excitation with momentum $k$ is therefore given by $(\delta\hat{B}_{k+q}^{\mathrm{c}})^\dagger(\delta\hat{B}^{\mathrm{c}}_{k-q})^\dag\ket{\beta^{\mathrm{c}}_0}$ and has energy $\mathcal{E}_{k+q}+\mathcal{E}_{k-q}$. 

Because of the gapless nature of the 1-particle Bogoliubov dispersion relation, the continuum of non-interacting 2-particle excitations is never separated in energy from the 1-particle dispersion, as seen in Figure~\ref{fig:Bogo-comparison} (right) and discussed in Appendix~\ref{app:Bogoliubov}. The gap between the isolated bound state (Goldstone mode) and the continuum of higher excitations is therefore a new feature of Gaussian TDVP due to the fact that it implements the interaction within the 1- and 2-particle sectors.

While the Goldstone mode continues to be well-described by the Bogoliubov dispersion relation $\mathcal{E}_k$, the spectrum of 2-particle excitations from Gaussian TDVP starts to divert as we increase $U$. In particular, we see that for sufficiently large $U$ both, the Goldstone mode and the doublon mode are completely separate from the continuum.

In figure~\ref{fig:Bogo-comparison}, we compare Gaussian TDVP and Bogoliubov theory in the regime where the Goldstone mode partially intersects with the continuum. We observe that this intersection appears for small $U$ in the Gaussian TDVP results only in those regimes where also the in Bogoliubov theory the 1-particle mode lies partially above the bottom of the many particle continuum. This phenomenon occurs for choices of $\mu$ and system dimension $\dim$ such that the dispersion relation $\mathcal{E}_k$ is not convex, \ie there exist $q,k$, such that $\mathcal{E}_k+\mathcal{E}_q<\mathcal{E}_{q+k}$. In Appendix~\ref{app:Bogoliubov}, we show that this can only happen for $\mu<6-2\dim$. The Gaussian TDVP continuum (light orange) agrees well with the Bogoliubov 2-particle spectrum (light red), where it intersects with the Goldstone mode, \ie roughly for $2\pi k/N\in (\pi/4,3\pi/4)$. Outside of this region, the two disagree: While the Gaussian TDVP gives rise to a finite gap between continuum and isolated Goldstone mode, 2-particle continuum and 1-particle dispersion relation necessarily touch for the non-interacting mean field Hamiltonian from Bogoliubov theory. 

In this regime of intersecting continuum and Goldstone mode, Gaussian TDVP can (at least partially) describe the decay of the Goldstone mode into the continuum of modes with a higher number of excitations. This phenomenon is known as Beliaev damping~\cite{beliaev-energy-1958} and is not captured by the standard Bogoliubov theory, but so far has been typically obtained from perturbative expansions by re-including higher order terms of the Hamiltonian. We will investigate this decay of 1-particle excitations into the continuum in section~\ref{sec:realtime}.

\subsection{Higgs mode}
Another suggestive observation can be made on the physical interpretation of the gapped mode at the bottom of the continuum. A possible interpretation is that it is a remnant of what, near the superfluid to Mott insulator transition, becomes known as the Higgs mode. It corresponds to oscillations of the amplitude of the order parameter $\braket{b_0}$ (while the Goldstone mode is interpreted as oscillations of the order parameter phase) and it has been observed experimentally by coupling to it through modulation of the tunneling amplitude~\cite{endres2012higgs}.

\begin{figure}[t!]
\centering
  \begin{tikzpicture}
  \draw (0,0) node[inner sep=0pt]{\includegraphics[width=\linewidth]{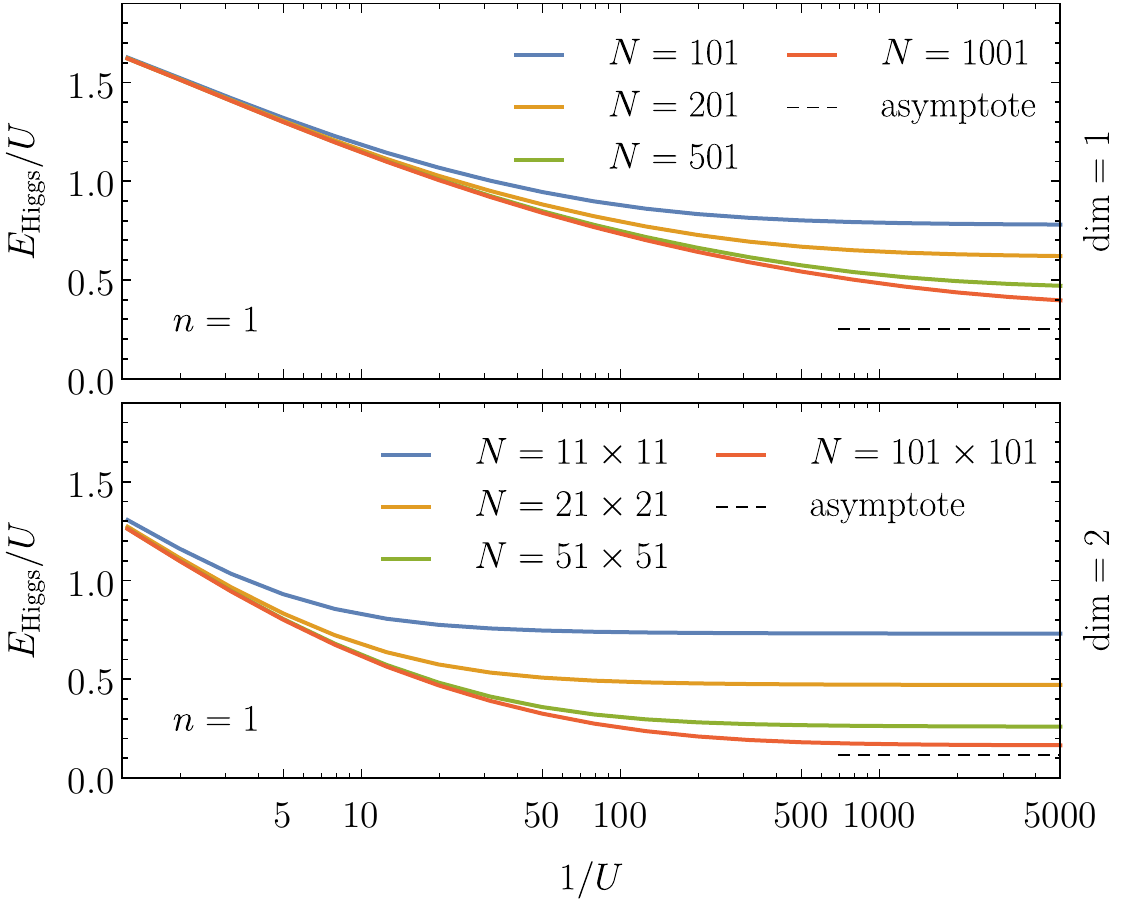}};
  \end{tikzpicture}
  \ccaption{Continuum/Higgs gap as function of $1/U$}{This figure shows the gap between the gapless Goldstone mode and the continuum of excitations as a function of $1/U$ in 1d and 2d and for different system sizes. The asymptotic value for large $N$ and small $U$ obtained in equation~\eqref{eq:higgsasymptotics} is also indicated for $N=1001$ in 1d and $N=101^2$ in 2d.}
  \label{fig:Higgs_gap}
\end{figure}

The prediction of our model for the continuum gap, which because of this possible interpretation we will label as $E_\mathrm{Higgs}$, can be studied numerically through the diagonalization of the matrix $K$ described in the previous paragraphs. At fixed non-zero $U$, $E_\mathrm{Higgs}$ converges to a finite non-zero value in the thermodynamic limit. 

\begin{figure*}[t!]
\centering
  \begin{tikzpicture}
  \draw (0,0) node[inner sep=0pt]{\includegraphics[width=\linewidth]{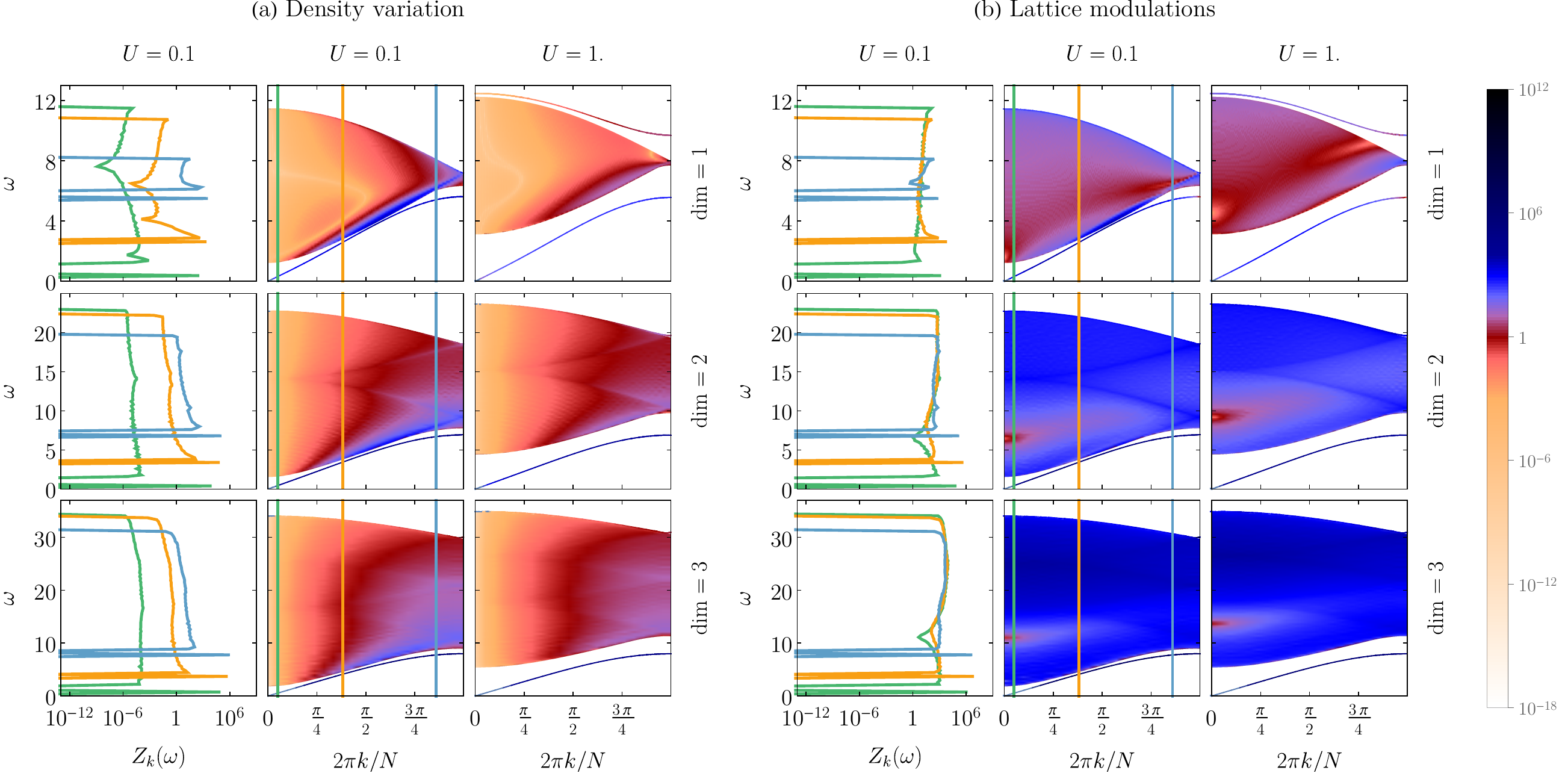}};
  \end{tikzpicture}
  \ccaption{Spectral functions for (a) Density variation $\hat{V}_{\mathrm{density}}^{(k)}$ from~\eqref{eq:V_density} and (b) Lattice modulation $\hat{V}_{\mathrm{lattice}}^{(k)}$ from~\eqref{eq:V_hopping}}{We show, as colour plots, the values of the spectral response functions $Z_k(\omega)$ in the relevant range of values of $k$ and $\omega$. In the first column of each panel, we show more logarithmic graphs of $Z_k(\omega)$ for fixed slice of $k$ (indicated by vertical lines of the respective color in the second column). The computations were performed for $N=(501,101^2,41^3)$ for $\dim=(1,2,3)$ respectively. To extract a continuous response functions, we performed a binning in energy intervals of $\Delta \omega=(0.13,0.2,0.37)$ for $\dim=(1,2,3)$ respectively.}
  \label{fig:SpectralFunctionDensity}
\end{figure*}

We are also able to give an analytical asymptotic result for the limit in which $U\to 0$ while $\mu$ varies so as to keep a constant particle number density of the ground state $n=\braket{\hat{N}}/N$ (see Appendix \ref{app:linearized_eom}). In this limit, we have that the gap goes to zero linearly in the interaction strength $U$, namely
\begin{align}
    \lim_{U\to 0}\frac{E_{\mathrm{Higgs}}}{U}=\alpha(N,n)&\sim 2 \sqrt[3]{2} n^{\frac{2}{3}} N^{-\frac{1}{3}}\quad\text{as}\quad N\to\infty\,.
    \label{eq:higgsasymptotics}
\end{align}
Note that it is instrumental that we took here first the limit $U\to 0$, before studying the large $N$ asymptotics.

In Figure~\ref{fig:Higgs_gap} one can see the numerical results for the behaviour of the ratio between the Higgs gap and $U$ and notice how it indeed approaches a constant asymptotic value for small $U$. In the large $U$ region, it has instead an unexpected divergent behaviour (the gap should close at the SF/MI transition~\cite{huber_dynamical_properties_2007}), however this can be understood as a breaking down of our model at the transition where Gaussian states are no longer a good description of the system's ground state. It is instead interesting to see how the constant small $U$ behaviour matches the experimentally measured value of the Higgs mode gap~\cite{endres2012higgs} even better than the previous theoretical results obtained with Gutzwiller theory.

\section{Linear Response}\label{sec:linear_response}
We use our variational manifold and the real time evolution projected onto it to study the response of the system to small perturbations. This is significant as it provides possible connections to actual experiments, where certain system responses can be probed and measured.

\begin{figure*}[htb]
\centering
  \begin{tikzpicture}
  \draw (0,0) node[inner sep=0pt]{\includegraphics[width=\linewidth]{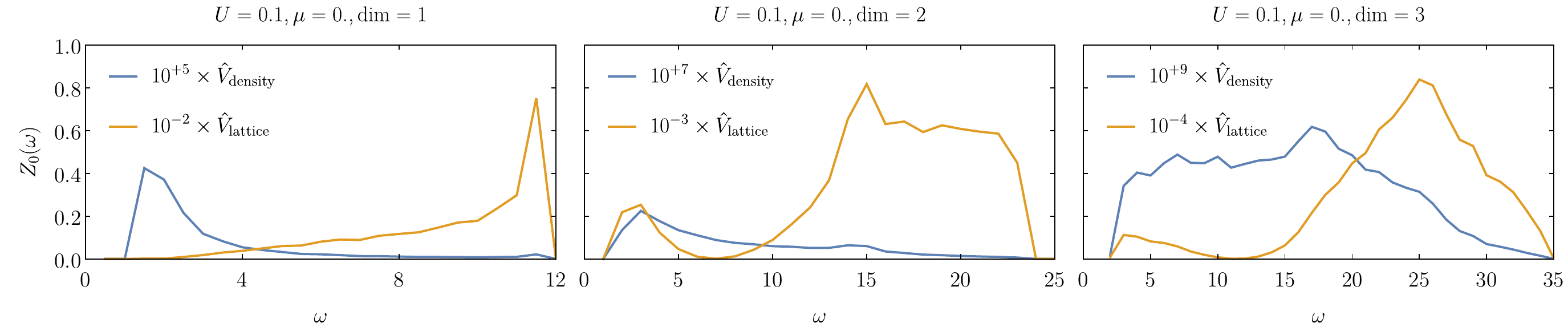}};
  \end{tikzpicture}
  \ccaption{Comparison: Density variation~\eqref{eq:V_density} and lattice modulation~\eqref{eq:V_hopping} at $k=0$}{We compare the response functions from density variation with the one from lattice modulations. The response to the lattice modulation is by several orders of magnitude stronger. Note that we rescaled the data by factors of $10^{\pm x}$ to fit into the same range. The computations were performed for $N=(501,101^2,41^3)$ for $\dim=(1,2,3)$ respectively. To extract a continuous response functions, we performed a binning in energy intervals of $\Delta \omega=(0.5,1,1)$ for $\dim=(1,2,3)$ respectively.}
  \label{fig:comparison_latticemodulation_vs_density}
\end{figure*}

\subsection{Spectral Functions}
We model an external perturbation by considering the time dependent perturbed Hamiltonian
\begin{align}
    \hat{H}_\lambda(t)=\hat{H}+\lambda \, \varphi(t) \, \hat{V}\,,
\end{align}
where $\hat{H}$ is the unperturbed Bose-Hubbard Hamiltonian~\eqref{eq:BH_Hamiltonian}, $\varphi(t)$ is a classical external field that couples to the system through the Hermitian operator $\hat{V}$ and $\lambda$ is a real parameter. We shall then consider the projected real time evolution $\ket{\psi_\lambda (t)}$ of the system under such perturbed Hamiltonian and evaluate its response in terms of the expectation value of the same coupling operator $\hat{V}$. In particular, we consider this response in the limit of small perturbations, \ie we compute quantities only up to first order in the parameter $\lambda$. Thus, we consider the response
\begin{align}
    \delta V(t)= {\frac{d}{d\lambda} \braket{\psi_\lambda(t)|\hat{V}|\psi_\lambda(t)}}_{\lambda=0}
\end{align}
to the perturbation $\hat{V}$.

As discussed in Appendix~\ref{app:linear_response_theory}, the Fourier space response $\delta V(\omega)$, calculated on the variational manifold as explained above, takes the form $\delta V(\omega)=\tilde{\varphi}(\omega) \, \chi(\omega)$, where $\tilde{\varphi}$ is the Fourier transform of the perturbing field $\varphi(t)$ and $Z_V(\omega)\equiv -\pi \mbox{Im}\chi(\omega)$ is the response function of the system with respect to the perturbation $\hat{V}$. Such response functions are expressed in terms of the spectral decomposition of the linearized real time evolution $K$ defined in Section \ref{sec:quasiparticle_excitations} as
\begin{align}
    Z_V(\omega)=\frac{1}{2}\,\mbox{sign}(\omega) \,  {\left|e^a(\omega) \, dV_a \right|}^2 \, \delta(|\omega|)\,,
    \label{eq:response_function}
\end{align}
where $dV$ is the gradient differential form of the real valued function on the manifold $\braket{\psi(\beta,h)|\hat{V}|\psi(\beta,h)}$, $e(\omega)$ are the eigenvectors of $K$ (defined in Appendix \ref{app:linearized_eom}) with eigenvalue $\omega$ and $\delta(|\omega|)$ is a normalization of the eigenvectors defined in equation~\eqref{eq:delta}.

\begin{figure}[t]
\centering
  \begin{tikzpicture}
  \draw (0,0) node[inner sep=0pt]{\includegraphics[width=\linewidth]{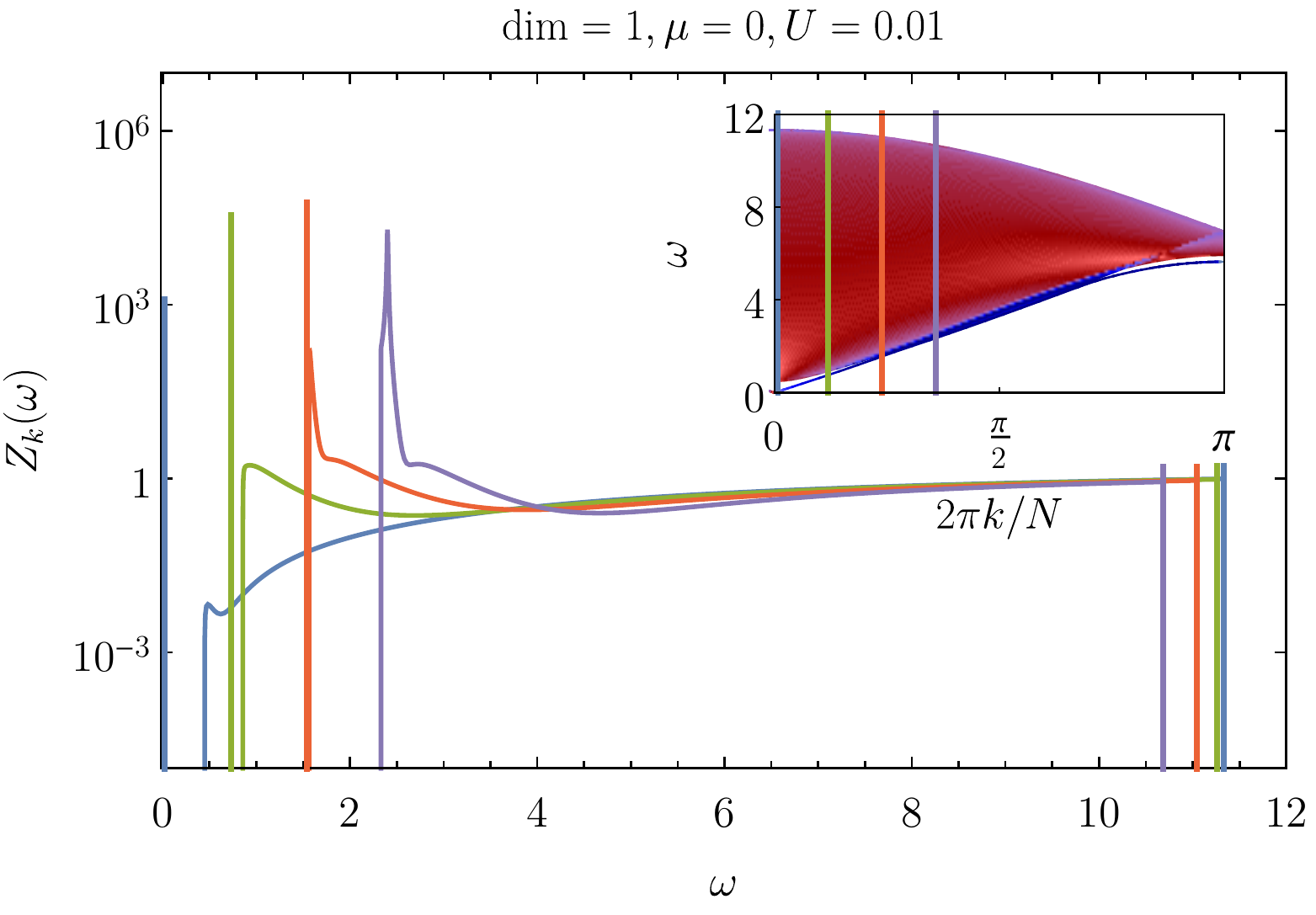}};
  \end{tikzpicture}
  \ccaption{Peak merging into continuum}{We plot the spectral response function $Z_k(\omega)$ relative to a density perturbation for a set of different momenta $k$. The function is calculated for a set of parameters ($\mbox{dim}=1$, $U=0.01$ and $\mu=0$) such that for some values of $k$ the Goldstone mode merges into the continuum spectrum. The plot shows how the delta-like peak of the Goldstone mode transforms into a finite width feature when this merge occurs (purple line).}
  \label{fig:SpectralMerging}
\end{figure}

For the Bose-Hubbard model, we consider the following types of perturbations:
\begin{align}
    \hat{V}^{(k)}_{\mathrm{1-particle}}&=\sum_k \mathcal{U}(x^{\mathrm{g}})(\ii\hat{b}_k^\dag - \ii\hat{b}_k) \,\mathcal{U}^\dag(x^{\mathrm{g}})\,.\label{eq:V_1-particle}\\
    \hat{V}^{(k)}_{\mathrm{density}}&=\sum_i \hat{b}_i^\dag \hat{b}_i \,\cos(k x_i)\label{eq:V_density}\,\\
    \hat{V}^{(k)}_{\mathrm{lattice}}&=\sum_{\langle i,j\rangle} (b_i^\dag \hat{b}_{j}+b_{j}^\dag \hat{b}_{i}) \,\cos(k x_i)\,\label{eq:V_hopping}\,.
\end{align}
In~\eqref{eq:V_1-particle}, we use a linear operator to create a single particle perturbation of momentum $k$. The other two perturbations are quadratic in creation and annihilation operators, such that the excitation consists in general of both single- and 2-particle excitations. In~\eqref{eq:V_density}, we consider a spatial \emph{density variation} by modulating the chemical potential with momentum $k$, which couples directly to the local particle density. In~\eqref{eq:V_hopping}, we consider a spatial modulation of momentum $k$ of the hopping constant. This can be achieved through a modulation of the lattice depth~\cite{kollath_spectroscopy_2006}. Such perturbation  naturally couples to the kinetic energy operator.

The different response functions $Z_k(\omega)$, obtained by evaluating~\eqref{eq:response_function} for different types of perturbation operators of momentum $k$ and at energy $\omega$, give us an indication of how strongly each type of perturbation couples to different regions of the spectrum.

A first observation we can make is on the behaviour of the isolated Goldstone mode in those situations when it merges with the continuum part of the spectrum. In Figure~\ref{fig:SpectralMerging}, we see how the isolated peak of the response function broadens into a wider feature inside the continuum. This indicates how, even when the Goldstone mode is not an isolated eigenstate, it still survives as a finite lifetime excitation of the system.

We can then also compare how the different perturbations considered couple to the system. In Figure \ref{fig:comparison_latticemodulation_vs_density}, we see how the perturbation that couples the strongest to the continuum modes at $k=0$ is the lattice modulation operator. 
Although it has to be mentioned that the definition of the normalizations of the perturbations~(\ref{eq:V_density}-\ref{eq:V_hopping}) is not free of some arbitrariness, the large difference in these coupling strengths provides a further element of support to the identification of the lower continuum modes as the Higgs excitation. Indeed it is known that the Higgs mode should be excited most easily through perturbations of the kinetic energy term of the Hamiltonian, while the Goldstone mode through perturbations in the particle density \cite{huber_dynamical_properties_2007}.

\subsection{Real time evolution}\label{sec:realtime}
\begin{figure}[t!]
\begin{tikzpicture}
\draw (0,0) node[inner sep=0pt]{\includegraphics[width=\linewidth]{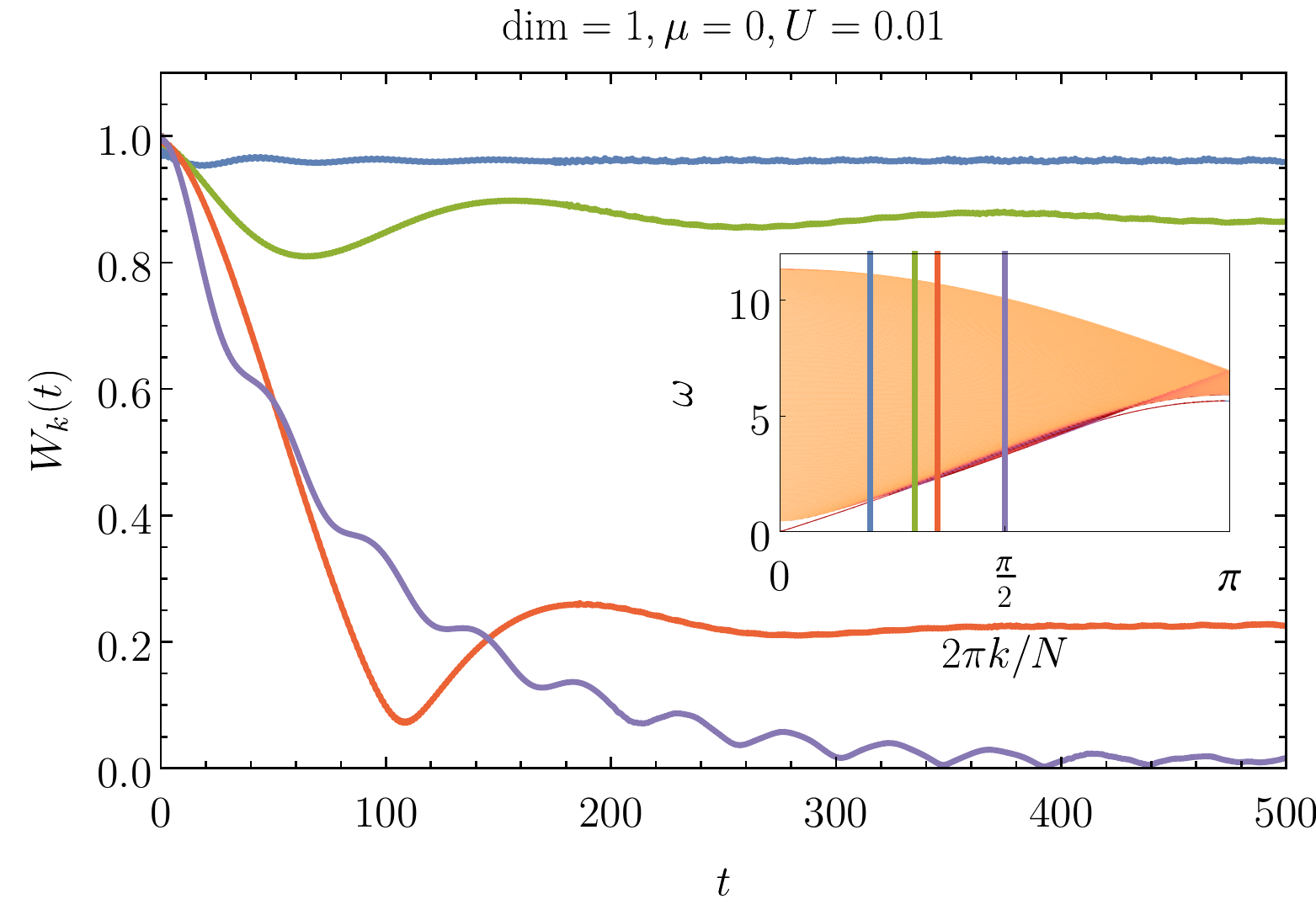}};
\end{tikzpicture}
\ccaption{Time evolution of 1-particle weight}{We show, for different momenta $k$, the real time evolution of the 1-particle weight $W_k(t)={|\braket{\psi_{k}(0)|\psi_{k}(t)}|}^2$, where $\ket{\psi_{k}}=\hat{V}^{(k)}_{\small{\text{1 particle}}}\ket{\psi_{\mathrm{g}}}$ is a perturbation vector in the one particle sector of the tangent plane. The overlap in the previous equation is computed as explained in Appendix~\ref{app:linear_response_theory}.}\label{fig:time_evolution}
\end{figure}

The analysis of the response function can also give indications on the real time evolution of perturbations of the system. Indeed, we can interpret the operator $\hat{V}$ as creating a perturbation described by the tangent vector $\ket{\delta\psi_{\mathrm{g}}(0)}=\mathbb{P}_{\ket{\psi_{\mathrm{g}}}}(-\ii \hat{V}) \ket{\psi_{\mathrm{g}}}$ at $t=0$, which is equivalent to giving the system a kick by choosing $\varphi(t)=\delta(t)$. The evolution in the tangent plane of this perturbation vector is then given by $d\Phi_t$ at $\ket{\psi_{\mathrm{g}}}$, \ie the push-forward of the real time evolution flow $\Phi_t$ around the stationary point. $d\Phi_t$ is a linear map on the tangent space at $\ket{\psi_{\mathrm{g}}}$, explicitly it is given by the matrix $e^{Kt}$ with respect to the basis~\eqref{eq:tangent_plane_basis}.

We consider an initial perturbation $\ket{\delta\psi_{\mathrm{g}}(0)}$ in the 1-particle sector of the tangent plane, \ie we require $\ket{\delta\psi_{\mathrm{g}}(0)}$ to be only spanned by 1-particle states in basis~\eqref{eq:tangent_plane_basis}. In particular, this is accomplished by the perturbation created by $\hat{V}^{(k)}_{1-\mathrm{particle}}$. The time evolution of $\ket{\delta\psi_{\mathrm{g}}(0)}$ under the map $d\Phi_t$ can then show different characteristic behaviour. If the perturbation is created at a momentum value where there exists an isolated Goldstone state with a strong coupling to the 1-particle sector (quantified by the 1-particle response function), the perturbation will persist indefinitely. If, instead, the excitation has a momentum at which the 1-particle perturbation couples sufficiently strongly to the continuum, a part of it will decay into the continuum modes, disappearing in a time proportional to the inverse of the width of the response function. Finally, if there is no isolated Goldstone state at the chosen momentum, but only the continuum, the perturbation will have a finite lifetime and decay completely into continuum excitations.

In Figure~\ref{fig:time_evolution}, we show the overlap of the time evolved perturbation with the 1-particle sector of the tangent plane, for different total momenta of the initial perturbation. A perturbation with momentum $k$ corresponding to an isolated Goldstone state will maintain a large overlap with the 1-particle sector. For perturbations with momentum $k$ closer to the region where the Goldstone mode merges with the continuum, a larger part of the overlap with the 1-particle sector will decay in time. Finally, if the perturbation has a total momentum $k$, for which no isolated Goldstone state exists in the spectrum, the single particle overlap will decay completely to zero after a finite lifetime.
Such decay behavior is similar to what can also be seen in quantum optical systems coupled to unconventional photon baths~\cite{gonzalez_markovian_2017}.

This behaviour of the evolution of perturbations can be interpreted as a remnant in lattice systems of what in continuum Bose-Einstein condensates (BEC) is known as the Beliaev damping of excitations, \ie the decay of 1-particle excitations into the continuum of many particle excitations due to scattering interactions. Our variational scheme successfully captures at least part of this behaviour, namely the one associated to the 1- and 2-particle sector that are fully included in our tangent space. This is in contrast to the traditional Bogoliubov theory that is restricted to the non-interacting 1-particle sector. In particular, standard Bogoliubov theory cannot describe the interaction with the continuum consistently, which can only be incorporated by re-including the previously neglected terms as perturbations~\cite{beliaev-energy-1958}.

\section{Relations between methods}\label{sec:comparison}
Our study is based on the time dependent variational principle (TDVP), where we project the equations of motion on a given variational class and linearize them around the stationary state that provides the best approximation of the ground state. While we focused on the class of all Gaussian states, the method can be applied to any suitable family of states, so it is natural to compare the results between different variational classes. In the context of Bogoliubov theory, it is natural to compare our larger manifold of \emph{all} bosonic Gaussian states $\mathcal{D}(\beta)\mathcal{S}(\lambda)\ket{0}$ with the smaller sub manifold consisting only of coherent (or displaced) states $\mathcal{D}(\beta)\ket{0}$. Table~\ref{tab:review_methods} summarizes the different methods.

\begin{table*}[t!]
	\centering
	\ccaption{Comparison of tangent plane methods}{
	We relate known methods to compute excitation spectra based on choosing a tangent plane of (i) coherent states around $\ket{\beta_0^{\mathrm{c}}}$, (ii) coherent states around $\ket{\psi_{\mathrm{g}}}$ and (iii) general Gaussian states around $\ket{\psi_{\mathrm{g}}}$ and list for which combinations, we find the expected gapless Goldstone mode.}
	\renewcommand{\arraystretch}{1.7}
	\setlength{\tabcolsep}{1mm}
	\begin{center}
	\begin{tabular}{c||l c l||l}
		 & \multicolumn{1}{c}{\textbf{Linearized TDVP}} & & & \multicolumn{1}{c}{\textbf{Projected Hamiltonian}} \\[-1.5mm]
		 & \multicolumn{1}{c}{spectrum of $K^a{}_b$} & & equivalent method: & \multicolumn{1}{c}{spectrum of $\mathbb{P}_{\ket{\psi}}\hat{H}\mathbb{P}_{\ket{\psi}}$} \\[1.5mm]
		\hline
		\hline
		\textbf{(i) Coherent states} & \multirow{2}{*}{gapless Goldstone mode $\mathcal{E}_k^{\mathrm{c}}$}  & \multirow{2}{*}{$\Rightarrow$} & Bogoliubov theory & \multirow{2}{*}{gapped 1-particle spectrum} \\[-1.5mm]
		around $\ket{\beta^{\mathrm{c}}_0}$ & & & (see Appendix~\ref{app:Bogoliubov}-\ref{app:Bogoliubov_spectrum_from_linearization}) & \\
		\hline
		\textbf{(ii) Coherent states} & \multirow{2}{*}{gapped 1-particle spectrum $\mathcal{E}_k^{\mathrm{g}}$}  & \multirow{2}{*}{$\Rightarrow$} & Iterated Bogoliubov theory & \multirow{2}{*}{gapped 1-particle spectrum} \\[-1.5mm]
		around $\ket{\psi_{\mathrm{g}}}$ & & & (see Appendix~\ref{app:iterated_Bogoliubov}) & \\
		\hline
		\textbf{(iii) Gaussian states } & gapless Goldstone mode, & \multirow{2}{*}{$\Rightarrow$} & Random phase approximation & gapped 1-particle spectrum,\\[-1.5mm]
		around $\ket{\psi_{\mathrm{g}}}$ & gapped 2-particle spectrum & & (see Appendix~\ref{app:feynman_diagrams}) & gapped 2-particle spectrum
	\end{tabular}
	\end{center}
	\label{tab:review_methods}
\end{table*}

\noindent\textbf{(i) Coherent TDVP around $\ket{\beta_0^{\mathrm{c}}}$.} If we apply linearized TDVP to the manifold of coherent states, we obtain the same excitation spectrum as the single-particle spectrum of Bogoliubov mean field theory (see Appendix~\ref{app:Bogoliubov_spectrum_from_linearization}). The latter is defined by taking the full Hamiltonian and using the commutation relations to normal order the creation and annihilation operators with respect to the coherent state $\ket{\beta_0^{\mathrm{c}}}$ that minimizes the energy on the coherent state manifold. At this point, we can truncate at quadratic order to obtain the mean field Hamiltonian $[\hat{H}]_{\ket{\beta_0^{\mathrm{c}}}}$ and use a Bogoliubov transformation to compute its excitation spectrum $\mathcal{E}_k$. However, we should point out that $[\hat{H}]_{\ket{\beta_0^{\mathrm{c}}}}$ contains more information than the linearized TDVP, as it gives us a Hamiltonian operator whose minimal energy $E_{\mathrm{Bogoliubov}}$ is a better estimate of the system's ground state energy than just $E_{\ket{\beta_0^{\mathrm{c}}}}$. On the other hand, this energy is not variational, \ie it cannot be expressed as the expectation value of an ansatz state on the full system Hamiltonian. Furthermore, the truncation of $[\hat{H}]_{\ket{\beta_0^{\mathrm{c}}}}$ is not self-consistent, because $\ket{\beta_0^{\mathrm{c}}}$ is not its ground state.

\noindent\textbf{(ii) Coherent TDVP around $\ket{\psi_0^{\mathrm{c}}}$.} After finding the best Gaussian ground state approximation $\ket{\psi_{\mathrm{g}}}$, we can linearize the equations of motion restricted to the space of displacements. This is equivalent to iterating traditional Bogoliubov theory as reviewed in appendix~\ref{app:iterated_Bogoliubov}, where we find the self-consistent mean field Hamiltonian $[\hat{H}]_{\ket{\psi_{\mathrm{g}}}}$, whose ground state is again $\ket{\psi_{\mathrm{g}}}$. The 1-particle spectrum $\mathcal{E}_k^{\mathrm{g}}$ from this Hamiltonian is gapped and consequently not a good approximation to the Goldstone mode. However, we can use $\mathcal{E}_k^{\mathrm{g}}$ to construct the 2-particle continuum of the quadratic Hamiltonian $[\hat{H}]_{\ket{\psi_{\mathrm{g}}}}$. Interestingly, the resulting 2-particle spectrum provides a good approximation to the continuum with Gaussian TDVP (see figure~\ref{fig:IteratedBogoliubovSpectrum}). In this way, we can understand the gap $\mathcal{E}^{\mathrm{g}}_0$ as already encoding the interaction energies between two particle excitations that is required to approximate the interacting 2-particle spectrum.

\noindent\textbf{(iii) Gaussian TDVP around $\ket{\beta_0^{\mathrm{c}}}$.} In order to obtain a self-consistent ground state, we  enlarge the manifold of states and introduce general Gaussian states, which also allow for squeezing. Indeed, the Gaussian state of minimal energy $\ket{\psi_{\mathrm{g}}}$ can also be identified as the state that fulfills the property of being the ground state of the corresponding mean field Hamiltonian $[\hat{H}]_{\ket{\psi_{\mathrm{g}}}}$, \ie the quadratic truncation of the full Hamiltonian when normal-ordered with respect to $\ket{\psi_{\mathrm{g}}}$. If we apply linearized TDVP to the extended manifold of Gaussian states, we obtain the spectrum object of this paper, which naturally contains both 1- and 2-particle excitations (see Section~\ref{sec:quasiparticle_excitations}). The Gaussian TDVP spectrum can be equivalently obtained using random phase approximation (see Appendix~\ref{app:feynman_diagrams}). More precisely, the Gaussian state $\ket{\psi_{\mathrm{g}}}$ can be used as the reference vacuum when expanding Green's functions in terms of Feynman diagrams. We can consistently resum all ladder diagrams to obtain an approximate excitation spectrum, that agrees with the one found through Gaussian TDVP.

\noindent\textbf{Projected Hamiltonian.} Finally, there is a well-known alternative~\cite{haegeman_postmps_2013,vanderstraeten_simulating_excitations_2019,shi_variational_2018} to compute excitation spectra from a tangent plane based on the projected Hamiltonian. Instead of linearizing the equations of motion, we can directly take the tangent plane as variational ansatz for eigenstates by projecting the full Hamiltonian onto it, \ie $H_{\mathbb{P}}=\mathbb{P}_{\ket{\psi}}\hat{H}\mathbb{P}_{\ket{\psi}}$, and then computing its spectrum. The eigenstates $\ket{E_i}$ with energy $E_i$ of the projected Hamiltonian $H_{\mathbb{P}}$ are manifestly variational, \ie their expectation value with respect to the full Hamiltonian is equal to $E_i$ and there exists a true eigenstate of the full Hamiltonian with smaller energy. This is not necessarily the case for the eigenvectors of $K^a{}_b$ in the linearized TDVP. In~\cite{haegeman_postmps_2013}, it has been further pointed out that--in contrast to the projected Hamiltonian method--the linearized TDVP may incorrectly predict massless excitation modes. This occurs whenever the approximate ground state within the chosen variational family spontaneously breaks a symmetry which is not spontaneously broken in the exact ground state. In the case of the Bose-Hubbard model, this is actually a desirable feature: while the true ground state only breaks the $\mathbb{U}(1)$ symmetry in the limit $N\to\infty$, the family of Gaussian states already breaks this symmetry for finite $N$ and is thus well-suited to study the superfluid phase in the thermodynamic limit.

\section{Discussion and outlook}\label{sec:discussion}
Our Gaussian TDVP method naturally generalizes Bogoliubov theory to describe the superfluid phase of the Bose-Hubbard model. The presented methods provide systematic framework to compute (a) approximate ground state energies, (b) excitation spectra and (c) linear response functions for general variational families.

(a) Our variational ansatz of all Gaussian states provided a good variational approximation of the degenerate set of ground states. While its ground energy prediction is generally worse than the ground state energy of the traditional Bogoliubov mean field Hamiltonian, our method has the advantage of being manifestly variational, \ie we find a concrete state whose energy expectation value with respect to the full Hamiltonian provides a rigorous upper bound to the true ground state energy.

(b) The linearized equations of motion projected onto the full Gaussian tangent plane allowed us to capture interactions between 1-particle excitations. We find a gapless mode that can be identified with the Goldstone mode that is well-approximated by traditional Bogoliubov theory. We point out that, compared to other variational approaches such as~\cite{knap_variational_2011}, our variational family is chosen so as to yield a Goldstone mode that is exactly gapless. We also find a gap between Goldstone mode and 2-particle continuum that is not captured by traditional Bogoliubov theory's non-interacting $n$-particle spectrum. We argued that the lowest band of the 2-particle continuum can be identified as the Higgs mode. Finally, for some parameter choices, we found an isolated doublon state.

(c) Using eigenvalues and eigenvectors of the linearized equations of motion, we could compute spectral functions for linear perturbations that are generated by arbitrary linear and quadratic operators. Here, we observed how these different perturbations coupled to isolated bound states, \eg the Goldstone and the doublon mode. Finally, we described the time evolution of the Goldstone mode, in particular its (partial) decay depending on its interplay with the 2-particle continuum.

An interesting feature of the TDVP is the interplay and relations to other methods, such as Bogoliubov theory and random phase approximation. For the computation of excitation spectra, we find that these standard methods give the same results as the TDVP computation for the correct choice of variational manifold. Another interesting observation is the fact that the 2-particle continuum of the Gaussian TDVP is well-approximated by the free 2-particle spectrum from coherent TDVP (or equivalently: iterated Bogoliubov theory) around the best Gaussian state $\ket{\psi_{\mathrm{g}}}$. This suggests that the 1- and 2-particle spectrum can be approximated by a hybrid approach with two different quadratic Hamiltonians, \ie the Bogoliubov Hamiltonian $[\hat{H}]_{\ket{\beta_0^{\mathrm{c}}}}$ and the iterated Bogoliubov Hamiltonian $[\hat{H}]_{\ket{\psi_{\mathrm{g}}}}$. While the former describes the massless Goldstone mode, we can use the latter's free 2-particle spectrum to approximate the interacting 2-particle spectrum of the full Hamiltonian.

The one of our predictions that calls most for further inquiry is the gapped $2$-particle continuum above the Goldstone mode. It will be interesting to further explore with other methods whether the identification of the lowest continuum mode as the Higgs mode is correct and whether the gap that separates it from the Goldstone mode survives, once one also considers excitations of three or more particles.

The presented scheme is self-consistent and requires no other assumptions than the choice of variational manifold. In particular, it can be easily applied to other variational families, such as non-Gaussian states~\cite{shi_variational_2018} or Gutzwiller states. To study the Bose-Hubbard model also in the Mott phase, we expect that a variational manifold that combines both, Gutzwiller states and Gaussian transformations, \ie the family of states resulting from applying a Gaussian unitary to a Gutzwiller product state, is particularly promising and will be analyzed in future work.

\begin{acknowledgements}
We thank Abhay Ashtekar, Bertrand Halperin, Andy Martin, Marcos Rigol, Richard Schmidt and Yao Wang for inspiring discussions. TG, LH and IC thank Harvard University for the hospitality during several visits. TG, LH, CH and IC are supported by the Deutsche Forschungsgemeinschaft (DFG, German Research Foundation) under Germany’s Excellence Strategy – EXC-2111 – 39081486. LH is funded by the the Max Planck Harvard Research Center for Quantum Optics. TS acknowledges the Thousand-Youth-Talent Program of China. CH and IC acknowledge funding through ERC Grant QUENOCOBA, ERC-2016-ADG (Grant no.742102). ED acknowledges funding through Harvard-MIT CUA, AFOSR-MURI: Photonic Quantum Matter (award FA95501610323) and DARPA DRINQS program (award D18AC00014).
\end{acknowledgements}

\bibliography{references}

\begin{thebibliography}{28}%
\makeatletter
\providecommand \@ifxundefined [1]{%
 \@ifx{#1\undefined}
}%
\providecommand \@ifnum [1]{%
 \ifnum #1\expandafter \@firstoftwo
 \else \expandafter \@secondoftwo
 \fi
}%
\providecommand \@ifx [1]{%
 \ifx #1\expandafter \@firstoftwo
 \else \expandafter \@secondoftwo
 \fi
}%
\providecommand \natexlab [1]{#1}%
\providecommand \enquote  [1]{``#1''}%
\providecommand \bibnamefont  [1]{#1}%
\providecommand \bibfnamefont [1]{#1}%
\providecommand \citenamefont [1]{#1}%
\providecommand \href@noop [0]{\@secondoftwo}%
\providecommand \href [0]{\begingroup \@sanitize@url \@href}%
\providecommand \@href[1]{\@@startlink{#1}\@@href}%
\providecommand \@@href[1]{\endgroup#1\@@endlink}%
\providecommand \@sanitize@url [0]{\catcode `\\12\catcode `\$12\catcode
  `\&12\catcode `\#12\catcode `\^12\catcode `\_12\catcode `\%12\relax}%
\providecommand \@@startlink[1]{}%
\providecommand \@@endlink[0]{}%
\providecommand \url  [0]{\begingroup\@sanitize@url \@url }%
\providecommand \@url [1]{\endgroup\@href {#1}{\urlprefix }}%
\providecommand \urlprefix  [0]{URL }%
\providecommand \Eprint [0]{\href }%
\providecommand \doibase [0]{http://dx.doi.org/}%
\providecommand \selectlanguage [0]{\@gobble}%
\providecommand \bibinfo  [0]{\@secondoftwo}%
\providecommand \bibfield  [0]{\@secondoftwo}%
\providecommand \translation [1]{[#1]}%
\providecommand \BibitemOpen [0]{}%
\providecommand \bibitemStop [0]{}%
\providecommand \bibitemNoStop [0]{.\EOS\space}%
\providecommand \EOS [0]{\spacefactor3000\relax}%
\providecommand \BibitemShut  [1]{\csname bibitem#1\endcsname}%
\let\auto@bib@innerbib\@empty
\bibitem [{\citenamefont {Bloch}\ \emph {et~al.}(2008)\citenamefont {Bloch},
  \citenamefont {Dalibard},\ and\ \citenamefont
  {Zwerger}}]{bloch_ultracold_gases_2008}%
  \BibitemOpen
  \bibfield  {author} {\bibinfo {author} {\bibfnamefont {I.}~\bibnamefont
  {Bloch}}, \bibinfo {author} {\bibfnamefont {J.}~\bibnamefont {Dalibard}}, \
  and\ \bibinfo {author} {\bibfnamefont {W.}~\bibnamefont {Zwerger}},\ }\href
  {\doibase 10.1103/RevModPhys.80.885} {\bibfield  {journal} {\bibinfo
  {journal} {Rev. Mod. Phys.}\ }\textbf {\bibinfo {volume} {80}},\ \bibinfo
  {pages} {885} (\bibinfo {year} {2008})}\BibitemShut {NoStop}%
\bibitem [{\citenamefont {Bogolyubov}(1947)}]{Bogolyubov_1947}%
  \BibitemOpen
  \bibfield  {author} {\bibinfo {author} {\bibfnamefont {N.~N.}\ \bibnamefont
  {Bogolyubov}},\ }\href@noop {} {\bibfield  {journal} {\bibinfo  {journal} {J.
  Phys.(USSR)}\ }\textbf {\bibinfo {volume} {11}},\ \bibinfo {pages} {23}
  (\bibinfo {year} {1947})},\ \bibinfo {note} {[Izv. Akad. Nauk Ser.
  Fiz.11,77(1947)]}\BibitemShut {NoStop}%
\bibitem [{\citenamefont {Pekker}\ \emph {et~al.}(2012)\citenamefont {Pekker},
  \citenamefont {Wunsch}, \citenamefont {Kitagawa}, \citenamefont {Manousakis},
  \citenamefont {S\o{}rensen},\ and\ \citenamefont
  {Demler}}]{pekker_signatures_2012}%
  \BibitemOpen
  \bibfield  {author} {\bibinfo {author} {\bibfnamefont {D.}~\bibnamefont
  {Pekker}}, \bibinfo {author} {\bibfnamefont {B.}~\bibnamefont {Wunsch}},
  \bibinfo {author} {\bibfnamefont {T.}~\bibnamefont {Kitagawa}}, \bibinfo
  {author} {\bibfnamefont {E.}~\bibnamefont {Manousakis}}, \bibinfo {author}
  {\bibfnamefont {A.~S.}\ \bibnamefont {S\o{}rensen}}, \ and\ \bibinfo {author}
  {\bibfnamefont {E.}~\bibnamefont {Demler}},\ }\href {\doibase
  10.1103/PhysRevB.86.144527} {\bibfield  {journal} {\bibinfo  {journal} {Phys.
  Rev. B}\ }\textbf {\bibinfo {volume} {86}},\ \bibinfo {pages} {144527}
  (\bibinfo {year} {2012})}\BibitemShut {NoStop}%
\bibitem [{\citenamefont {Sachdev}(1999)}]{sachdev_quantum_1999}%
  \BibitemOpen
  \bibfield  {author} {\bibinfo {author} {\bibfnamefont {S.}~\bibnamefont
  {Sachdev}},\ }\href@noop {} {\emph {\bibinfo {title} {Quantum {Phase}
  {Transitions}}}}\ (\bibinfo  {publisher} {Cambridge University Press},\
  \bibinfo {address} {Cambridge, UK},\ \bibinfo {year} {1999})\BibitemShut
  {NoStop}%
\bibitem [{\citenamefont {Huber}\ \emph {et~al.}(2008)\citenamefont {Huber},
  \citenamefont {Theiler}, \citenamefont {Altman},\ and\ \citenamefont
  {Blatter}}]{huber_amplitude_mode_2008}%
  \BibitemOpen
  \bibfield  {author} {\bibinfo {author} {\bibfnamefont {S.~D.}\ \bibnamefont
  {Huber}}, \bibinfo {author} {\bibfnamefont {B.}~\bibnamefont {Theiler}},
  \bibinfo {author} {\bibfnamefont {E.}~\bibnamefont {Altman}}, \ and\ \bibinfo
  {author} {\bibfnamefont {G.}~\bibnamefont {Blatter}},\ }\href {\doibase
  10.1103/PhysRevLett.100.050404} {\bibfield  {journal} {\bibinfo  {journal}
  {Phys. Rev. Lett.}\ }\textbf {\bibinfo {volume} {100}},\ \bibinfo {pages}
  {050404} (\bibinfo {year} {2008})}\BibitemShut {NoStop}%
\bibitem [{\citenamefont {Huber}\ \emph {et~al.}(2007)\citenamefont {Huber},
  \citenamefont {Altman}, \citenamefont {B\"uchler},\ and\ \citenamefont
  {Blatter}}]{huber_dynamical_properties_2007}%
  \BibitemOpen
  \bibfield  {author} {\bibinfo {author} {\bibfnamefont {S.~D.}\ \bibnamefont
  {Huber}}, \bibinfo {author} {\bibfnamefont {E.}~\bibnamefont {Altman}},
  \bibinfo {author} {\bibfnamefont {H.~P.}\ \bibnamefont {B\"uchler}}, \ and\
  \bibinfo {author} {\bibfnamefont {G.}~\bibnamefont {Blatter}},\ }\href
  {\doibase 10.1103/PhysRevB.75.085106} {\bibfield  {journal} {\bibinfo
  {journal} {Phys. Rev. B}\ }\textbf {\bibinfo {volume} {75}},\ \bibinfo
  {pages} {085106} (\bibinfo {year} {2007})}\BibitemShut {NoStop}%
\bibitem [{\citenamefont {Bissbort}\ \emph {et~al.}(2014)\citenamefont
  {Bissbort}, \citenamefont {Buchhold},\ and\ \citenamefont
  {Hofstetter}}]{bissbort_quasi-particle_2014}%
  \BibitemOpen
  \bibfield  {author} {\bibinfo {author} {\bibfnamefont {U.}~\bibnamefont
  {Bissbort}}, \bibinfo {author} {\bibfnamefont {M.}~\bibnamefont {Buchhold}},
  \ and\ \bibinfo {author} {\bibfnamefont {W.}~\bibnamefont {Hofstetter}},\
  }\href {http://arxiv.org/abs/1401.4466} {\bibfield  {journal} {\bibinfo
  {journal} {arXiv:1401.4466 [cond-mat, physics:quant-ph]}\ } (\bibinfo {year}
  {2014})}\BibitemShut {NoStop}%
\bibitem [{\citenamefont {Sengupta}\ and\ \citenamefont
  {Dupuis}(2005)}]{sengupta_mott-insulatorsuperfluid_2005}%
  \BibitemOpen
  \bibfield  {author} {\bibinfo {author} {\bibfnamefont {K.}~\bibnamefont
  {Sengupta}}\ and\ \bibinfo {author} {\bibfnamefont {N.}~\bibnamefont
  {Dupuis}},\ }\href {\doibase 10.1103/PhysRevA.71.033629} {\bibfield
  {journal} {\bibinfo  {journal} {Phys. Rev. A}\ }\textbf {\bibinfo {volume}
  {71}} (\bibinfo {year} {2005}),\ 10.1103/PhysRevA.71.033629}\BibitemShut
  {NoStop}%
\bibitem [{\citenamefont {Fitzpatrick}\ and\ \citenamefont
  {Kennett}(2018)}]{fitzpatrick_contour-time_2018}%
  \BibitemOpen
  \bibfield  {author} {\bibinfo {author} {\bibfnamefont {M.~R.~C.}\
  \bibnamefont {Fitzpatrick}}\ and\ \bibinfo {author} {\bibfnamefont {M.~P.}\
  \bibnamefont {Kennett}},\ }\href {\doibase
  https://doi.org/10.1016/j.nuclphysb.2018.02.021} {\bibfield  {journal}
  {\bibinfo  {journal} {Nuclear Physics B}\ }\textbf {\bibinfo {volume}
  {930}},\ \bibinfo {pages} {1 } (\bibinfo {year} {2018})}\BibitemShut
  {NoStop}%
\bibitem [{\citenamefont {Knap}\ \emph {et~al.}(2011)\citenamefont {Knap},
  \citenamefont {Arrigoni},\ and\ \citenamefont {von~der
  Linden}}]{knap_variational_2011}%
  \BibitemOpen
  \bibfield  {author} {\bibinfo {author} {\bibfnamefont {M.}~\bibnamefont
  {Knap}}, \bibinfo {author} {\bibfnamefont {E.}~\bibnamefont {Arrigoni}}, \
  and\ \bibinfo {author} {\bibfnamefont {W.}~\bibnamefont {von~der Linden}},\
  }\href {\doibase 10.1103/PhysRevB.83.134507} {\bibfield  {journal} {\bibinfo
  {journal} {Phys. Rev. B}\ }\textbf {\bibinfo {volume} {83}},\ \bibinfo
  {pages} {134507} (\bibinfo {year} {2011})}\BibitemShut {NoStop}%
\bibitem [{\citenamefont {Menotti}\ and\ \citenamefont
  {Trivedi}(2008)}]{menotti_spectral_weight_2008}%
  \BibitemOpen
  \bibfield  {author} {\bibinfo {author} {\bibfnamefont {C.}~\bibnamefont
  {Menotti}}\ and\ \bibinfo {author} {\bibfnamefont {N.}~\bibnamefont
  {Trivedi}},\ }\href {\doibase 10.1103/PhysRevB.77.235120} {\bibfield
  {journal} {\bibinfo  {journal} {Phys. Rev. B}\ }\textbf {\bibinfo {volume}
  {77}},\ \bibinfo {pages} {235120} (\bibinfo {year} {2008})}\BibitemShut
  {NoStop}%
\bibitem [{\citenamefont {Bijlsma}\ and\ \citenamefont
  {Stoof}(1997)}]{stoof_variational_1997}%
  \BibitemOpen
  \bibfield  {author} {\bibinfo {author} {\bibfnamefont {M.}~\bibnamefont
  {Bijlsma}}\ and\ \bibinfo {author} {\bibfnamefont {H.~T.~C.}\ \bibnamefont
  {Stoof}},\ }\href {\doibase 10.1103/PhysRevA.55.498} {\bibfield  {journal}
  {\bibinfo  {journal} {Phys. Rev. A}\ }\textbf {\bibinfo {volume} {55}},\
  \bibinfo {pages} {498} (\bibinfo {year} {1997})}\BibitemShut {NoStop}%
\bibitem [{\citenamefont {Endres}\ \emph {et~al.}(2012)\citenamefont {Endres},
  \citenamefont {Fukuhara}, \citenamefont {Pekker}, \citenamefont {Cheneau},
  \citenamefont {Schau{\ss}}, \citenamefont {Gross}, \citenamefont {Demler},
  \citenamefont {Kuhr},\ and\ \citenamefont {Bloch}}]{endres2012higgs}%
  \BibitemOpen
  \bibfield  {author} {\bibinfo {author} {\bibfnamefont {M.}~\bibnamefont
  {Endres}}, \bibinfo {author} {\bibfnamefont {T.}~\bibnamefont {Fukuhara}},
  \bibinfo {author} {\bibfnamefont {D.}~\bibnamefont {Pekker}}, \bibinfo
  {author} {\bibfnamefont {M.}~\bibnamefont {Cheneau}}, \bibinfo {author}
  {\bibfnamefont {P.}~\bibnamefont {Schau{\ss}}}, \bibinfo {author}
  {\bibfnamefont {C.}~\bibnamefont {Gross}}, \bibinfo {author} {\bibfnamefont
  {E.}~\bibnamefont {Demler}}, \bibinfo {author} {\bibfnamefont
  {S.}~\bibnamefont {Kuhr}}, \ and\ \bibinfo {author} {\bibfnamefont
  {I.}~\bibnamefont {Bloch}},\ }\href
  {https://www.nature.com/articles/nature11255} {\bibfield  {journal} {\bibinfo
   {journal} {Nature}\ }\textbf {\bibinfo {volume} {487}},\ \bibinfo {pages}
  {454} (\bibinfo {year} {2012})}\BibitemShut {NoStop}%
\bibitem [{\citenamefont {Bissbort}\ \emph {et~al.}(2011)\citenamefont
  {Bissbort}, \citenamefont {G\"{o}tze}, \citenamefont {Li}, \citenamefont
  {Heinze}, \citenamefont {Krauser}, \citenamefont {Weinberg}, \citenamefont
  {Becker}, \citenamefont {Sengstock},\ and\ \citenamefont
  {Hofstetter}}]{bissbort_detecting_2011}%
  \BibitemOpen
  \bibfield  {author} {\bibinfo {author} {\bibfnamefont {U.}~\bibnamefont
  {Bissbort}}, \bibinfo {author} {\bibfnamefont {S.}~\bibnamefont {G\"{o}tze}},
  \bibinfo {author} {\bibfnamefont {Y.}~\bibnamefont {Li}}, \bibinfo {author}
  {\bibfnamefont {J.}~\bibnamefont {Heinze}}, \bibinfo {author} {\bibfnamefont
  {J.~S.}\ \bibnamefont {Krauser}}, \bibinfo {author} {\bibfnamefont
  {M.}~\bibnamefont {Weinberg}}, \bibinfo {author} {\bibfnamefont
  {C.}~\bibnamefont {Becker}}, \bibinfo {author} {\bibfnamefont
  {K.}~\bibnamefont {Sengstock}}, \ and\ \bibinfo {author} {\bibfnamefont
  {W.}~\bibnamefont {Hofstetter}},\ }\href {\doibase
  10.1103/PhysRevLett.106.205303} {\bibfield  {journal} {\bibinfo  {journal}
  {Phys. Rev. Lett.}\ }\textbf {\bibinfo {volume} {106}},\ \bibinfo {pages}
  {205303} (\bibinfo {year} {2011})}\BibitemShut {NoStop}%
\bibitem [{\citenamefont {Schori}\ \emph {et~al.}(2004)\citenamefont {Schori},
  \citenamefont {St\"oferle}, \citenamefont {Moritz}, \citenamefont {K\"ohl},\
  and\ \citenamefont {Esslinger}}]{shori_excitations_2004}%
  \BibitemOpen
  \bibfield  {author} {\bibinfo {author} {\bibfnamefont {C.}~\bibnamefont
  {Schori}}, \bibinfo {author} {\bibfnamefont {T.}~\bibnamefont {St\"oferle}},
  \bibinfo {author} {\bibfnamefont {H.}~\bibnamefont {Moritz}}, \bibinfo
  {author} {\bibfnamefont {M.}~\bibnamefont {K\"ohl}}, \ and\ \bibinfo {author}
  {\bibfnamefont {T.}~\bibnamefont {Esslinger}},\ }\href {\doibase
  10.1103/PhysRevLett.93.240402} {\bibfield  {journal} {\bibinfo  {journal}
  {Phys. Rev. Lett.}\ }\textbf {\bibinfo {volume} {93}},\ \bibinfo {pages}
  {240402} (\bibinfo {year} {2004})}\BibitemShut {NoStop}%
\bibitem [{\citenamefont {Weedbrook}\ \emph {et~al.}(2012)\citenamefont
  {Weedbrook}, \citenamefont {Pirandola}, \citenamefont
  {Garc\'{i}a-Patr\'{o}n}, \citenamefont {Cerf}, \citenamefont {Ralph},
  \citenamefont {Shapiro},\ and\ \citenamefont
  {Lloyd}}]{weedbrook_gaussian_2012}%
  \BibitemOpen
  \bibfield  {author} {\bibinfo {author} {\bibfnamefont {C.}~\bibnamefont
  {Weedbrook}}, \bibinfo {author} {\bibfnamefont {S.}~\bibnamefont
  {Pirandola}}, \bibinfo {author} {\bibfnamefont {R.}~\bibnamefont
  {Garc\'{i}a-Patr\'{o}n}}, \bibinfo {author} {\bibfnamefont {N.~J.}\
  \bibnamefont {Cerf}}, \bibinfo {author} {\bibfnamefont {T.~C.}\ \bibnamefont
  {Ralph}}, \bibinfo {author} {\bibfnamefont {J.~H.}\ \bibnamefont {Shapiro}},
  \ and\ \bibinfo {author} {\bibfnamefont {S.}~\bibnamefont {Lloyd}},\ }\href
  {\doibase 10.1103/RevModPhys.84.621} {\bibfield  {journal} {\bibinfo
  {journal} {Rev. Mod. Phys.}\ }\textbf {\bibinfo {volume} {84}},\ \bibinfo
  {pages} {621} (\bibinfo {year} {2012})}\BibitemShut {NoStop}%
\bibitem [{\citenamefont {Beliaev}(1958)}]{beliaev-energy-1958}%
  \BibitemOpen
  \bibfield  {author} {\bibinfo {author} {\bibfnamefont {S.~T.}\ \bibnamefont
  {Beliaev}},\ }\href {http://www.jetp.ac.ru/cgi-bin/e/index/e/7/2/p299?a=list}
  {\bibfield  {journal} {\bibinfo  {journal} {JETP}\ }\textbf {\bibinfo
  {volume} {7}},\ \bibinfo {pages} {299} (\bibinfo {year} {1958})}\BibitemShut
  {NoStop}%
\bibitem [{\citenamefont {White}(1992)}]{white1992density}%
  \BibitemOpen
  \bibfield  {author} {\bibinfo {author} {\bibfnamefont {S.~R.}\ \bibnamefont
  {White}},\ }\href {\doibase 10.1103/PhysRevLett.69.2863} {\bibfield
  {journal} {\bibinfo  {journal} {Phys. Rev. Lett.}\ }\textbf {\bibinfo
  {volume} {69}},\ \bibinfo {pages} {2863} (\bibinfo {year}
  {1992})}\BibitemShut {NoStop}%
\bibitem [{\citenamefont {Haegeman}\ \emph {et~al.}(2011)\citenamefont
  {Haegeman}, \citenamefont {Cirac}, \citenamefont {Osborne}, \citenamefont
  {Pi\ifmmode~\check{z}\else \v{z}\fi{}orn}, \citenamefont {Verschelde},\ and\
  \citenamefont {Verstraete}}]{haegeman_tdvp_2011}%
  \BibitemOpen
  \bibfield  {author} {\bibinfo {author} {\bibfnamefont {J.}~\bibnamefont
  {Haegeman}}, \bibinfo {author} {\bibfnamefont {J.~I.}\ \bibnamefont {Cirac}},
  \bibinfo {author} {\bibfnamefont {T.~J.}\ \bibnamefont {Osborne}}, \bibinfo
  {author} {\bibfnamefont {I.}~\bibnamefont {Pi\ifmmode~\check{z}\else
  \v{z}\fi{}orn}}, \bibinfo {author} {\bibfnamefont {H.}~\bibnamefont
  {Verschelde}}, \ and\ \bibinfo {author} {\bibfnamefont {F.}~\bibnamefont
  {Verstraete}},\ }\href {\doibase 10.1103/PhysRevLett.107.070601} {\bibfield
  {journal} {\bibinfo  {journal} {Phys. Rev. Lett.}\ }\textbf {\bibinfo
  {volume} {107}},\ \bibinfo {pages} {070601} (\bibinfo {year}
  {2011})}\BibitemShut {NoStop}%
\bibitem [{\citenamefont {Dirac}(1930)}]{dirac_note_1930}%
  \BibitemOpen
  \bibfield  {author} {\bibinfo {author} {\bibfnamefont {P.~A.~M.}\
  \bibnamefont {Dirac}},\ }\href {\doibase 10.1017/S0305004100016108}
  {\bibfield  {journal} {\bibinfo  {journal} {Mathematical Proceedings of the
  Cambridge Philosophical Society}\ }\textbf {\bibinfo {volume} {26}},\
  \bibinfo {pages} {376–385} (\bibinfo {year} {1930})}\BibitemShut {NoStop}%
\bibitem [{\citenamefont {Pitaevskii}(1961)}]{pitaevskii_vortex_1961}%
  \BibitemOpen
  \bibfield  {author} {\bibinfo {author} {\bibfnamefont {L.~P.}\ \bibnamefont
  {Pitaevskii}},\ }\href
  {http://www.jetp.ac.ru/cgi-bin/e/index/e/13/2/p451?a=list} {\bibfield
  {journal} {\bibinfo  {journal} {JETP}\ }\textbf {\bibinfo {volume} {13}},\
  \bibinfo {pages} {451} (\bibinfo {year} {1961})}\BibitemShut {NoStop}%
\bibitem [{\citenamefont {Gross}(1961)}]{gross_structure_1961}%
  \BibitemOpen
  \bibfield  {author} {\bibinfo {author} {\bibfnamefont {E.~P.}\ \bibnamefont
  {Gross}},\ }\href {\doibase 10.1007/BF02731494} {\bibfield  {journal}
  {\bibinfo  {journal} {Il Nuovo Cimento (1955-1965)}\ }\textbf {\bibinfo
  {volume} {20}},\ \bibinfo {pages} {454} (\bibinfo {year} {1961})}\BibitemShut
  {NoStop}%
\bibitem [{\citenamefont {Kollath}\ \emph {et~al.}(2006)\citenamefont
  {Kollath}, \citenamefont {Iucci}, \citenamefont {Giamarchi}, \citenamefont
  {Hofstetter},\ and\ \citenamefont
  {Schollw\"ock}}]{kollath_spectroscopy_2006}%
  \BibitemOpen
  \bibfield  {author} {\bibinfo {author} {\bibfnamefont {C.}~\bibnamefont
  {Kollath}}, \bibinfo {author} {\bibfnamefont {A.}~\bibnamefont {Iucci}},
  \bibinfo {author} {\bibfnamefont {T.}~\bibnamefont {Giamarchi}}, \bibinfo
  {author} {\bibfnamefont {W.}~\bibnamefont {Hofstetter}}, \ and\ \bibinfo
  {author} {\bibfnamefont {U.}~\bibnamefont {Schollw\"ock}},\ }\href {\doibase
  10.1103/PhysRevLett.97.050402} {\bibfield  {journal} {\bibinfo  {journal}
  {Phys. Rev. Lett.}\ }\textbf {\bibinfo {volume} {97}},\ \bibinfo {pages}
  {050402} (\bibinfo {year} {2006})}\BibitemShut {NoStop}%
\bibitem [{\citenamefont {Gonz\'alez-Tudela}\ and\ \citenamefont
  {Cirac}(2017)}]{gonzalez_markovian_2017}%
  \BibitemOpen
  \bibfield  {author} {\bibinfo {author} {\bibfnamefont {A.}~\bibnamefont
  {Gonz\'alez-Tudela}}\ and\ \bibinfo {author} {\bibfnamefont {J.~I.}\
  \bibnamefont {Cirac}},\ }\href {\doibase 10.1103/PhysRevA.96.043811}
  {\bibfield  {journal} {\bibinfo  {journal} {Phys. Rev. A}\ }\textbf {\bibinfo
  {volume} {96}},\ \bibinfo {pages} {043811} (\bibinfo {year}
  {2017})}\BibitemShut {NoStop}%
\bibitem [{\citenamefont {Haegeman}\ \emph {et~al.}(2013)\citenamefont
  {Haegeman}, \citenamefont {Osborne},\ and\ \citenamefont
  {Verstraete}}]{haegeman_postmps_2013}%
  \BibitemOpen
  \bibfield  {author} {\bibinfo {author} {\bibfnamefont {J.}~\bibnamefont
  {Haegeman}}, \bibinfo {author} {\bibfnamefont {T.~J.}\ \bibnamefont
  {Osborne}}, \ and\ \bibinfo {author} {\bibfnamefont {F.}~\bibnamefont
  {Verstraete}},\ }\href {\doibase 10.1103/PhysRevB.88.075133} {\bibfield
  {journal} {\bibinfo  {journal} {Phys. Rev. B}\ }\textbf {\bibinfo {volume}
  {88}},\ \bibinfo {pages} {075133} (\bibinfo {year} {2013})}\BibitemShut
  {NoStop}%
\bibitem [{\citenamefont {Vanderstraeten}\ \emph {et~al.}(2019)\citenamefont
  {Vanderstraeten}, \citenamefont {Haegeman},\ and\ \citenamefont
  {Verstraete}}]{vanderstraeten_simulating_excitations_2019}%
  \BibitemOpen
  \bibfield  {author} {\bibinfo {author} {\bibfnamefont {L.}~\bibnamefont
  {Vanderstraeten}}, \bibinfo {author} {\bibfnamefont {J.}~\bibnamefont
  {Haegeman}}, \ and\ \bibinfo {author} {\bibfnamefont {F.}~\bibnamefont
  {Verstraete}},\ }\href {\doibase 10.1103/PhysRevB.99.165121} {\bibfield
  {journal} {\bibinfo  {journal} {Phys. Rev. B}\ }\textbf {\bibinfo {volume}
  {99}},\ \bibinfo {pages} {165121} (\bibinfo {year} {2019})}\BibitemShut
  {NoStop}%
\bibitem [{\citenamefont {Shi}\ \emph {et~al.}(2018)\citenamefont {Shi},
  \citenamefont {Demler},\ and\ \citenamefont {Cirac}}]{shi_variational_2018}%
  \BibitemOpen
  \bibfield  {author} {\bibinfo {author} {\bibfnamefont {T.}~\bibnamefont
  {Shi}}, \bibinfo {author} {\bibfnamefont {E.}~\bibnamefont {Demler}}, \ and\
  \bibinfo {author} {\bibfnamefont {J.~I.}\ \bibnamefont {Cirac}},\ }\href
  {\doibase 10.1016/j.aop.2017.11.014} {\bibfield  {journal} {\bibinfo
  {journal} {Annals of Physics}\ }\textbf {\bibinfo {volume} {390}},\ \bibinfo
  {pages} {245} (\bibinfo {year} {2018})}\BibitemShut {NoStop}%
\bibitem [{\citenamefont {Demler}\ \emph {et~al.}(1996)\citenamefont {Demler},
  \citenamefont {Zhang}, \citenamefont {Bulut},\ and\ \citenamefont
  {Scalapino}}]{demler_collective_excitations_1996}%
  \BibitemOpen
  \bibfield  {author} {\bibinfo {author} {\bibfnamefont {E.}~\bibnamefont
  {Demler}}, \bibinfo {author} {\bibfnamefont {S.~C.}\ \bibnamefont {Zhang}},
  \bibinfo {author} {\bibfnamefont {N.}~\bibnamefont {Bulut}}, \ and\ \bibinfo
  {author} {\bibfnamefont {D.~J.}\ \bibnamefont {Scalapino}},\ }\href {\doibase
  10.1142/S0217979296000982} {\bibfield  {journal} {\bibinfo  {journal}
  {International Journal of Modern Physics B}\ }\textbf {\bibinfo {volume}
  {10}},\ \bibinfo {pages} {2137} (\bibinfo {year} {1996})}\BibitemShut
  {NoStop}%
\end{thebibliography}%

\clearpage

\appendix
\section{Review of Bogoliubov theory}\label{app:Bogoliubov}
In the main part of this paper, we generalize the well-known Bogoliubov theory of the Bose-Hubbard model by extending the underlying variational family from coherent to squeezed coherent states. To allow for a fair comparison, we now review the three steps involved in traditional Bogoliubov theory.

For later computations, it is useful to write the Bose-Hubbard Hamiltonian~\eqref{eq:BH_Hamiltonian} in momentum space
\begin{align}
    \hat{H}=\sum_{k}\varepsilon_k\hat{b}_k^\dagger \hat{b}_k+\frac{U}{2N}\sum_{k,p,q}\hat{b}^\dagger_{k+q}\hat{b}^\dagger_{p-q}\hat{b}_{k}\hat{b}_p
\end{align}
where we defined $\hat{b}_k=\frac{1}{\sqrt{N}} \sum_i e^{-\ii k x_i} \hat{b}_i$ on the reciprocal lattice and introduced the non-interacting dispersion relation
\begin{align}
    \varepsilon_k=-2\sum^{\dim}_{d=1} \mbox{cos}\frac{2\pi k_d}{N_d} -\mu\,,
    \label{eq:nonint_epsilon}
\end{align}
where $N_d$ refers to the number of lattice sites in the $d$-th direction, such that $N=\prod_dN_d$.

\noindent\textbf{Step 1 (coherent variation).} Bogoliubov theory approximates the ground state within the class of translationally invariant coherent states, \ie the states 
\begin{align}
    \ket{\beta_0}=\mathcal{D}(\beta_0)\ket{0}\quad\text{with}\quad \mathcal{D}(\beta_0)=e^{\beta_0 \hat{b}_0^\dagger-\beta_0^*\hat{b}_0}\,,
\end{align}
satisfying $\bra{\beta_0}b_k\ket{\beta_0}=\beta_0 \delta_{0,k}$. Within this class, the average energy value is minimized for $|\beta_0|$ equal to
\begin{align}
    \beta^{\mathrm{c}}_0:=\sqrt{-\varepsilon_0 N/U}\label{eq:betac0}
\end{align}
leading to the expectation value
\begin{align}
    E_{\ket{\beta^{\mathrm{c}}_0}}=\bra{\beta^{\mathrm{c}}_0}\hat{H}\ket{\beta^{\mathrm{c}}_0}=\varepsilon_0\,|\beta^{\mathrm{c}}_0|^2+\frac{U}{2N}\,|\beta^{\mathrm{c}}_0|^4=-\frac{\epsilon_0^2\,N}{2U}\,,\label{eq:E_beta0}
\end{align}
provided that $\varepsilon_0<0$. There is a larger set of solutions given by $\beta_0=\beta_0^{\mathrm{c}}e^{\ii\varphi}$ associated to the spontaneously broken $\mathrm{U}(1)$ symmetry generated by $\hat{N}=\sum_k\hat{b}_k^\dagger \hat{b}_k$.

\noindent\textbf{Step 2 (mean field Hamiltonian).} We can use $\ket{\beta^{\mathrm{c}}_0}$ to define the mean field Hamiltonian
\begin{align}
    [\hat{H}]_{\ket{\beta^{\mathrm{c}}_0}}=E_{\ket{\beta^{\mathrm{c}}_0}}\!+\frac{1}{2}\!\sum_{k}\left(U_k^{\mathrm{c}}(\delta\hat{b}_k^{\mathrm{c}})^\dagger \delta\hat{b}_k^{\mathrm{c}}+V_k^{\mathrm{c}}\delta\hat{b}_k^{\mathrm{c}} \delta\hat{b}_{-k}^{\mathrm{c}}+\text{H.c.}\right),
\end{align}
where $\delta\hat{b}_k^{\mathrm{c}}=\hat{b}_k-\delta_{k,0}\beta^{\mathrm{c}}_0$, $U_k^{\mathrm{c}}=\varepsilon_k+\frac{2U}{N}\,|\beta^{\mathrm{c}}_0|^2=\varepsilon_k-2\varepsilon_0$ and $V_k^{\mathrm{c}}=\frac{U}{N}\,(\beta^{\mathrm{c}}_0)^2=-\varepsilon_0$. Here, we define $[\hat{H}]_{\ket{\psi}}$ to be the quadratic Hamiltonian resulting from the quadratic truncation of $\hat{H}$ written as normal ordered polynomial in creation and annihilation operators $\delta\hat{b}_k^\dagger$ and $\delta\hat{b}_k$ associated to the Gaussian state $\ket{\psi}$, \ie in our case, $\delta\hat{b}_k\ket{\beta^{\mathrm{c}}_0}=0$.

\noindent\textbf{Step 3 (squeezed ground state).} The mean field Hamiltonian $[\hat{H}]_{\ket{\beta^{\mathrm{c}}_0}}$ is quadratic, implying that we can diagonalize it by applying the Bogoliubov transformation
\begin{align}
    \textstyle \mathcal{S}(\lambda)=\exp\left(\frac{1}{2}\sum_k(\lambda_k\,\hat{b}_k^\dagger \hat{b}_{-k}^\dagger-\lambda_k^*\,\hat{b}_{k}\hat{b}_{-k})\right)\,.
\end{align}
We perform the transformation by expressing $\hat{b}_k$ in terms of new creation and annihilation operators
\begin{align}
    \delta\hat{B}_k&=\mathcal{D}(\beta)\mathcal{S}(\lambda)\,\hat{b}_k\,\mathcal{S}^\dagger(\lambda)\mathcal{D}^\dag(\beta)\\
    &=u_k(\hat{b}_k-\beta_k)-v_k(\hat{b}_{-k}^\dagger-\beta_k)
\end{align}
The diagonalization is accomplished by $\beta^{\mathrm{c}}_0$, $u_k^{\mathrm{c}}=\cosh{\lambda_k^{\mathrm{c}}}$ and $v_k^{\mathrm{c}}=\sinh{\lambda_k^{\mathrm{c}}}$, where $\lambda^{\mathrm{c}}_k$ is given by
\begin{align}
	    \tanh{2\lambda_k^{\mathrm{c}}}=-\frac{V_k^{\mathrm{c}}}{U_k^{\mathrm{c}}}\,,
\end{align}
such that the Hamiltonian takes the form
\begin{align}
    [\hat{H}]_{\ket{\beta^{\mathrm{c}}_0}}=E_{\ket{\beta^{\mathrm{c}}_0}}-\Delta^{\mathrm{c}}+\sum_{k}\mathcal{E}_k^{\mathrm{c}}\,(\delta \hat{B}_k^{\mathrm{c}})^\dagger \delta \hat{B}_k^{\mathrm{c}}\,,
\end{align}
where the energy shift $\Delta^{\mathrm{c}}$ and the excitations $\mathcal{E}_k^{\mathrm{c}}$ are
\begin{align}
    \Delta^{\mathrm{c}}&=\frac{1}{2}\sum_k\frac{\varepsilon_0^2}{\varepsilon_k-2\varepsilon_0+\sqrt{(2\varepsilon_0-\varepsilon_k)^2-\varepsilon_0^2}}\,,\\
    \mathcal{E}^{\mathrm{c}}_k&=\sqrt{(U_k^{\mathrm{c}})^2-(V^{\mathrm{c}}_k)^2}=\sqrt{(\varepsilon_k-2\varepsilon_0)^2-\varepsilon_0^2}\,.\label{eq:BogoEk}
\end{align}
The formal ground state of this Hamiltonian is given by a state $\ket{\beta_0^{\mathrm{c}},\lambda^{\mathrm{c}}}=\mathcal{D}(\beta_0^{\mathrm{c}})\mathcal{S}(\lambda^{\mathrm{c}})\ket{0}$, such that $\delta\hat{B}^{\mathrm{c}}_k\ket{\beta_0^{\mathrm{c}},\lambda^{\mathrm{c}}}=0$. The Bogoliubov spectrum is gapless, \ie we have $\mathcal{E}^{\mathrm{c}}_0=0$.

\begin{figure}[t]
\centering
  \begin{tikzpicture}
  \draw (0,0) node[inner sep=0pt]{\includegraphics[width=\linewidth]{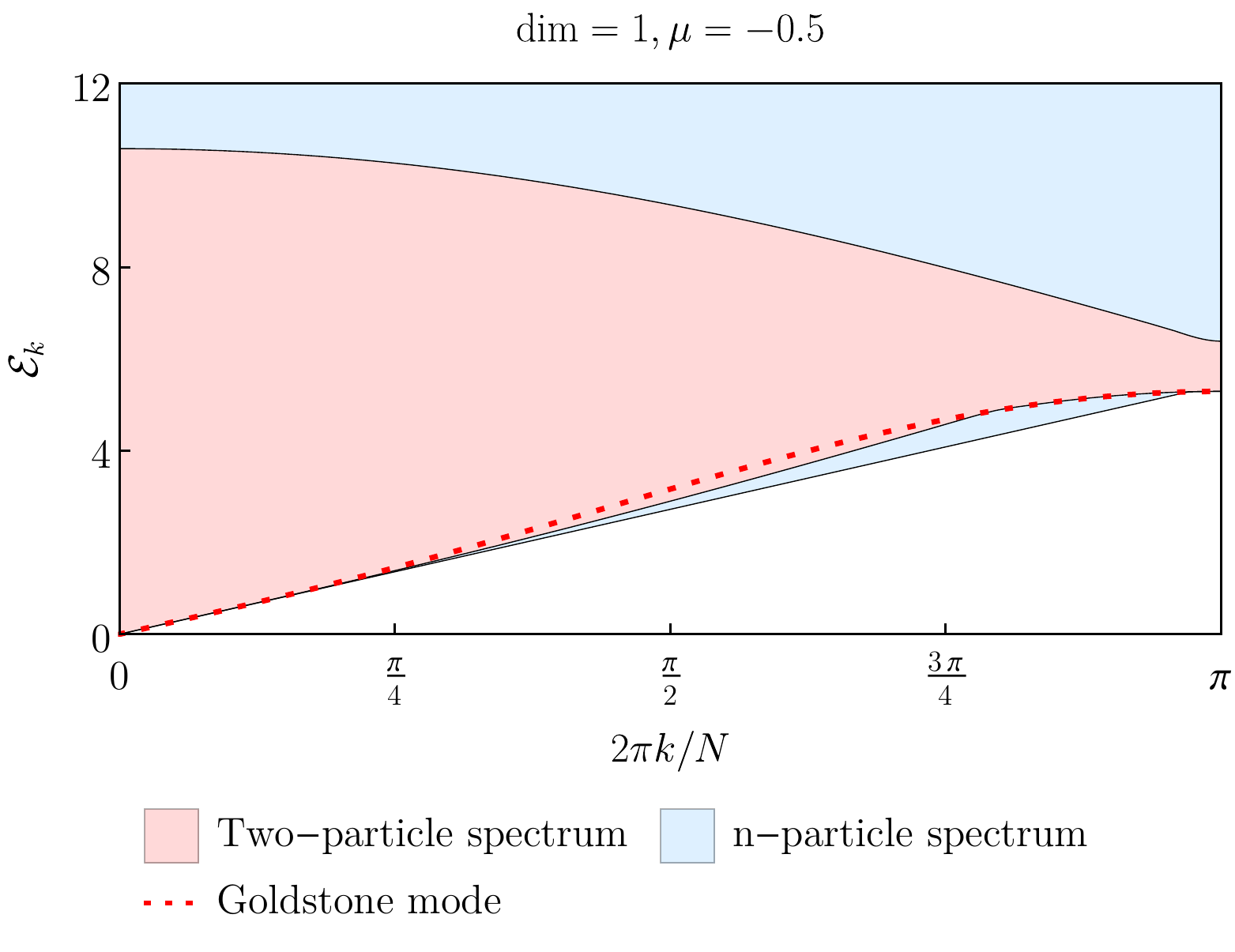}};
  \end{tikzpicture}
  \ccaption{Bogoliubov spectrum}{We show the Bogoliubov dispersion relation $\mathcal{E}_k$ as dashed line (red) within the 2-particle spectrum (light red) which is again embedded into the full excitation spectrum (light blue). Note that $\mathcal{E}_k$ only depends on $\mu$ and the dimension, but is independent of $U$.}
  \label{fig:Bogoliubov_spectrum}
\end{figure}

The full spectrum of the mean field Hamiltonian $[\hat{H}]_{\ket{\beta^{\mathrm{c}}_0}}$ is constructed from the 1-particle Bogoliubov spectrum $\mathcal{E}^{\mathrm{c}}_k$, whose excitations are non-interacting. Depending on the shape of the function $\mathcal{E}^{\mathrm{c}}_k$, \ie if there exist $k,q$ with $\mathcal{E}^{\mathrm{c}}_{k}>\mathcal{E}^{\mathrm{c}}_{k-q}+\mathcal{E}_{q}^{\mathrm{c}}$, it is possible to find a superposition of excitations with total momentum $k$ whose total energy is less than $\mathcal{E}_k$. To get the lower bound on the excitation continuum, we need to compute
\begin{align}
    \mathcal{E}_k^{\min}=\min_{q_i}\sum_{i}\mathcal{E}^{\mathrm{c}}_{q_i}\quad\text{with}\quad k=\sum_iq_i\,.
\end{align}
For the system in one dimension, it is sufficient to compute the slope for $k=0$, namely $\mathcal{E}'_0$, to find the explicit form of the lower bound to be given by
\begin{align}
    \mathcal{E}_k^{\min}=\min\left(\mathcal{E}_k,\sqrt{4+2\mu}\,\frac{2\pi k}{N}\right)\,.
\end{align}
We can derive the condition on $\mu$, such that there is some range of $k$, for which $\mathcal{E}_k^{\min}$ lies underneath the Bogoliubov spectrum $\mathcal{E}^{\mathrm{c}}_k$. This condition is given by $\partial^2_{k}\mathcal{E}^\mathrm{c}_k(\mu)=0$, \ie the second derivative of $\mathcal{E}_k^{\mathrm{c}}$ must vanish. Its solution is
\begin{align}
    \frac{2\pi k}{N}=\cos^{-1}\left(\frac{8+3 \mu-\sqrt{5 \mu ^2+32 \mu +48}}{4}\right)\,,
\end{align}
which only exists for $\mu\leq 4$. In higher dimensions, we can consider the slice $k=(k_x,0,\dots,0)$, which leads to an effective rescaling of $\mu\to \mu+2(\dim-1)$. In this case, we therefore have the condition $\mu< 6-2\dim$ to have part of the continuum spectrum to lie underneath the 1-particle dispersion relation $\mathcal{E}_k$.\\

Let us make the following three important remarks. First, the Bogoliubov energy
\begin{align}
    E_{\mathrm{Bogoliubov}}=E_{\ket{\beta^{\mathrm{c}}_0}}-\Delta^{\mathrm{c}}\label{eq:EBogo}
\end{align}
is not variational, \ie it is not the expectation value of the state $\ket{\beta_0^{\mathrm{c}},\lambda^{\mathrm{c}}}$ with respect to the full Hamiltonian, but rather the minimal energy of the mean field Hamiltonian $[\hat{H}]_{\ket{\beta^{\mathrm{c}}_0}}$. Only the energy $E_{\ket{\beta^{\mathrm{c}}_0}}$ is variational, \ie it minimizes the energy expectation value within the class of coherent states.\\
Second, the state $\ket{\beta_0^{\mathrm{c}},\lambda^{\mathrm{c}}}$ is actually ill defined in the zero mode, due to $\lambda_k^{\mathrm{c}}\to\infty$ for $k\to 0$. Put differently, the minimal energy $E_{\mathrm{Bogoliubov}}$ is only reached in the limit of an infinitely squeezed state, whose energy with respect to the full Hamiltonian actually diverges.\\
Third, we could have computed the Bogoliubov dispersion relation without defining the mean field Hamiltonian $[\hat{H}]_{\ket{\beta^{\mathrm{c}}_0}}$, but rather just by studying the real time flow of the full Hamiltonian projected on the manifold of coherent states and linearized around the stationary coherent state $\ket{\beta^{\mathrm{c}}_0}$ (see Appendix \ref{app:Bogoliubov_spectrum_from_linearization}). Thus, we do not need to perform the second and third step, if we are only interested in the variational ground state energy $E_{\ket{\beta^{\mathrm{c}}_0}}$ and the dispersion relation $\mathcal{E}^{\mathrm{c}}_k$, \ie we are content with not computing $\Delta$. In this case, it is sufficient to linearize Hamiltonian flow around the variational state $\ket{\beta^{\mathrm{c}}_0}$ and compute the spectrum of its generator. This is exactly what we do in this paper, but for the extended variational family of \emph{all} Gaussian states.

\section{Bogoliubov theory as coherent TDVP}\label{app:Bogoliubov_spectrum_from_linearization}
In this section, we show explicitly that coherent TDVP gives rise to the same 1-particle spectrum as Bogoliubov theory. We consider the manifold of displaced vacua (coherent states)
\begin{align}
    \ket{\beta}=\mathcal{D}(\beta)\ket{0}\quad\text{with}\quad \mathcal{D}(\beta)=e^{\sum_k \left(\beta_k \hat{b}_k^\dagger-\beta_k^*\hat{b}_k\right)}\,.
    \label{eq:displaced_states}
\end{align}
Here, $\beta$ is a vector written in the momentum basis, \ie its components are labeled by $k$. The tangent plane at the state $\ket{\beta}$ is spanned by vectors of the form $\mathcal{D}(\beta)\hat{b}_k^\dag \ket{0}$. Therefore, the projected real time evolution can be computed from the quantity
\begin{align}
    h_k(\beta)=\braket{0|\hat{b}_k \mathcal{D}^\dag(\beta) \hat{H}\mathcal{D}(\beta)|0}\,,
\end{align}
which for the Bose Hubbard model evaluates to
\begin{align}
    h_k(\beta)=\varepsilon_k\beta_k + \frac{U}{N}\sum_{k_1,k_2} \beta^*_{k_1+k_2-k} \beta_{k_1} \beta_{k_2}\,.
\end{align}
Expressed in real components, the resulting evolution is
\begin{align}
    \left(\begin{array}{c} \mbox{Re}\dot{\beta} \\ \mbox{Im}\dot{\beta} \end{array}\right)=\left(\begin{array}{c} \mbox{Im} \, h(\beta) \\ -\mbox{Re} \, h(\beta) \end{array}\right)=-\ii \frac{1}{\sqrt{2}}T^{-1}\,\left(\begin{array}{c} h(\beta) \\-h^*(\beta) \end{array}\right)\,,
\end{align}
where we introduced the transformation matrix
\begin{align}
    T=\frac{1}{\sqrt{2}}\begin{pmatrix}
    \id & \ii\id \\
    \id & -\ii\id \end{pmatrix}\,. \label{eq:matrix T}
\end{align}
The symbols $\beta$ and $h$ denote the column vectors that group the values of $\beta_k$ and $h_k$ for all values of $k$.

The linearization around $\ket{\beta^{\mathrm{c}}_0}$ is then given by
\begin{align}
    K&=\left(\begin{array}{cc} \frac{\partial}{\partial\mathrm{Re}\beta}, & \frac{\partial}{\partial\mathrm{Im}\beta} \end{array}\right) \left(\begin{array}{c} \mathrm{Re}\dot{\beta} \\ \mathrm{Im}\dot{\beta} \end{array}\right)\\
    &= \sqrt{2} \left(\begin{array}{cc} \frac{\partial}{\partial\beta}, & \frac{\partial}{\partial\beta^*} \end{array}\right) \left(\begin{array}{c} \mathrm{Re}\dot{\beta} \\ \mathrm{Im}\dot{\beta} \end{array}\right)  T \nonumber\\
    &=-\ii\, T^{-1}\left(\begin{array}{cc} \frac{\partial}{\partial\beta}, & \frac{\partial}{\partial\beta^*} \end{array}\right) \left(\begin{array}{c} h(\beta) \\-h^*(\beta) \end{array}\right)  T \,,
\end{align}
where we expressed the derivatives with respect to $\mathrm{Re}\beta$ and $\mathrm{Im}\beta$ in terms of derivatives with respect to $\beta$ and $\beta^*$, taken as independent variables. All derivatives are evaluated at $\beta_k=\delta_{k,0}\beta^{\mathrm{c}}_0$.
The matrix $\ii K$, whose eigenvalues $\pm\omega$ represent the TDVP estimate of the 1-particle excitation energies of the model, is then, up to similarity transformations, equal to
\begin{align}
    K={\left[\left(\begin{array}{cc} \frac{\partial}{\partial\beta}, & \frac{\partial}{\partial\beta^*} \end{array}\right) \left(\begin{array}{c} h(\beta) \\-h^*(\beta) \end{array}\right)\right]}_{\beta_k=\delta_{k,0}\beta^{\mathrm{c}}_0}.
\end{align}
This matrix decomposes into blocks of the form
\begin{align}
    K_k=\left(\begin{array}{cc} U_k^{\mathrm{c}} &  V^{\mathrm{c}}_k \\ -V^{\mathrm{c}}_k & -U^{\mathrm{c}}_k\end{array}\right)
\end{align}
with $U^{\mathrm{c}}_k=\varepsilon_k-2\varepsilon_0$ and $V^{\mathrm{c}}_k=-\varepsilon_0$ as in Appendix~\ref{app:Bogoliubov}. The eigenvalues $\pm\omega_k$ are given by
\begin{align}
    \omega_k=\sqrt{(U^{\mathrm{c}}_k)^2-(V^{\mathrm{c}}_k)^2}=\mathcal{E}^{\mathrm{c}}_k\,,
\end{align}
which is in full agreement with Bogoliubov theory~\eqref{eq:BogoEk}.

\section{Iterated Bogoliubov theory}\label{app:iterated_Bogoliubov}
Bogoliubov theory is not self-consistent, in the sense that we construct the mean field Hamiltonian $[\hat{H}]_{\ket{\beta^{\mathrm{c}}_0}}$ from the displaced state $\ket{\beta^{\mathrm{c}}_0}$, but when we compute the ground state of $[\hat{H}]_{\ket{\beta^{\mathrm{c}}_0}}$, we do not find $\ket{\beta^{\mathrm{c}}_0}$ again. We can therefore ask if there exists a state $\ket{\psi}$, which is the ground state of its own mean field Hamiltonian $[\hat{H}]_{\ket{\psi}}$.

A natural way to find this state consists of applying the Bogoliubov procedure repeatedly:

\noindent\textbf{Step 1 (displacement).}	Starting with the Gaussian state $\ket{\beta_0,\lambda}$ with real $\beta_0$ and $\lambda$, we can choose a new real displacement $\beta_0'$, such that the energy is minimized. The resulting $\beta_0'$ is given by
	\begin{align}
		\beta'_0=\sqrt{-\frac{N\varepsilon_0}{U}-\sum_k\left(2v_k^2+v_ku_k\right)}\,,
	\end{align}
where $u_k=\cosh{\lambda}$ and $v_k=\sinh{\lambda}$.

\noindent\textbf{Step 2 (quadratic expansion).} Once the new displaced ground state $\ket{\beta_0',\lambda}$ has been found, we can compute the unique quadratic Hamiltonian $[\hat{H}]_{\ket{\beta_0',\lambda}}$ through the following steps: First, we express the Hamiltonian in terms of the new annihilation operators
	\begin{align}
	    \delta\hat{b}'_k=u_k\,(\hat{b}_k-\delta_{0,k}\,\beta'_0)-v_{k}\,(\hat{b}_{-k}^\dagger-\delta_{0,k}\,\beta'_0)\,,
	\end{align}
that annihilate $\ket{\beta'_0,\lambda}$. Second, we use the canonical commutation relations $[\delta\hat{b}'_k,\delta\hat{b}'_{p}{}^{\dagger}]=\delta_{k,p}$ to write the Hamiltonian as sum of normal ordered operators. Third and finally, we truncate the resulting Hamiltonian at quadratic order to define
	\begin{align}
	[\hat{H}]_{\ket{\beta_0,\lambda}}=E_{\ket{\beta_0,\lambda}}\!+\frac{1}{2}\!\sum_{k}\left(\tilde{U}'_k\delta\hat{b}'_k{}^{\hspace{-.5mm}\dagger} \delta\hat{b}'_k+\tilde{V}'_k\delta\hat{b}'_k{} \delta\hat{b}'_{-k}{}+\text{H.c.}\right).
	\end{align}
Note that this is the straight-forward generalization of how we constructed the mean field Hamiltonian $[\hat{H}]_{\ket{\beta_0^{\mathrm{c}}}}$ in Bogoliubov theory. The coefficients $\tilde{U}_k$ and $\tilde{V}_k$ are explicitly given by
	\begin{align}
	\begin{split}
	    \tilde{U}_k&=\varepsilon_k\left(u_k^2+v_k^2\right)+\frac{2U}{N}u_kv_k\Big(\sum_qu_qv_q+\beta'_0{}^2\Big)\\
	    &\quad+\frac{2U}{N}\Big(u_k^2+v_q^2\Big)\Big(\sum_qv_q^2+\beta_0^2\Big)\,,
	\end{split}\\
	\begin{split}
	    \tilde{V}_k&=2\varepsilon_ku_kv_k+\frac{U}{N}\Big(u_k^2+v_k^2\Big)\Big(\sum_qu_qv_q+\beta'_0{}^2\Big)\\
	    &\quad+\frac{4U}{N}u_kv_k\Big(\sum_qv_q^2+\beta'_0{}^{2}\Big)\,.
	\end{split}
	\end{align}

\noindent\textbf{Step 3 (new ground state).} The ground state $\ket{\beta_0',\lambda'}$ of the mean field Hamiltonian can be encoded in another Bogoliubov transformation
	\begin{align}
	    \delta\hat{B}'_k{}=u_k'\,(\hat{b}_k-\delta_{0,k}\,\beta'_0)-v_k'\,(\hat{b}_{-k}^\dagger-\beta'_0\delta_{0,k})\,.
	\end{align}
Here, we have $u_k'=u_k\,\tilde{u}'_k+v_k\,\tilde{v}_k$ and $v_k'=u_k\tilde{v}_k+v_k\tilde{u}'_k$ with $\tilde{u}'_k=\cosh{\tilde{\lambda}'_k}$, $\tilde{v}_k=\sinh{\tilde{\lambda}'_k}$ and $\tanh{2\tilde{\lambda}'_k}=-\tilde{V}_k/\tilde{U}_k$. The ground state energy of the mean field Hamiltonian is given by
	\begin{align}
	    \bra{\beta_0',\lambda'}[\hat{H}]_{\ket{\beta'_0,\lambda}}\ket{\beta_0',\lambda'}=E_{\ket{\beta'_0,\lambda}}-\tilde{\Delta}'\,,
	\end{align}
with $\tilde{\Delta}'=-\sum_k\left(\tilde{U}_k\,(\tilde{v}_k')^2+\tilde{V}_k\,\tilde{u}_k'\tilde{v}_k'\right)$. The resulting excitation spectrum is given by
	\begin{align}
	    \mathcal{E}'_k=\sqrt{\tilde{U}_k^2-\tilde{V}_k^2}\,.\label{eq:E'_k}
	\end{align}
\begin{figure}
    \centering
    \begin{tikzpicture}[scale=1.3]
	\manifold[black,thick,fill=white,fill opacity=0.9]{0,-2.1}{4,-.8}
	\draw (4,-.8) node[above,yshift=7pt,xshift=-3pt]{$\vdots$};
	\draw (0,-2.1) node[above,yshift=7pt,xshift=3pt]{$\vdots$};
	\fill (2,-1.45) node[right]{$\ket{\psi_{\mathrm{g}}}$} circle (1.5pt);
	\draw[gray,thick,->] (2,-.55) -- (2,-1.45);
	\manifold[black,thick,fill=white,fill opacity=0.9]{0,-1.2}{4,.1}
	\fill (1.7, -.8) node[left,yshift=-1pt]{$\ket{\la{\lambda'},\be{\beta_0'}}$} circle (1.5pt);
	\draw[gray,thick,->] (1.7, -.1) -- (1.7, -.8);
	\manifold[black,thick,fill=white,fill opacity=0.9]{0,-.5}{4,.8}
	\fill (1.7, -.1) node[left,yshift=-1pt]{$\ket{\la{\lambda},\be{\beta_0'}}$} circle (1.5pt);
	\draw[gray,thick,->] (2.1, .1) -- (1.7, -.1);
	\fill (2.1, .1) node[right]{$\ket{\la{\lambda},\be{\beta_0^{\mathrm{c}}}}$} circle (1.5pt);
	\draw[gray,thick,->] (2.1, .8) -- (2.1, .1);
	\manifold[black,thick,fill=white,fill opacity=0.95]{0,0.2}{4,1.5}
	\fill (2.1, .8) node[right,yshift=1pt]{$\ket{\be{\beta_0^{\mathrm{c}}}}$} circle (1.5pt);
	\fill (1.5, .7) node[left,yshift=1pt]{$\ket{0}$} circle (1.5pt);
	\draw[gray,thick,->] (1.5, .7) -- (2.1, .8);
	\end{tikzpicture}
    \ccaption{Iterated Bogoliubov theory}{This figure illustrates the procedure of applying the steps of Bogoliubov theory iteratively to find best Gaussian approximation $\ket{\psi_{\mathrm{g}}}$ to the ground state of the Bose-Hubbard model. The individual sheets represent coherent states, \ie states with fixed $\lambda$, but different values of $\beta$. Traditional Bogoliubov is based on the first two steps from $\ket{0}$ to $\ket{\beta_0^{\mathrm{c}}}$ to $\ket{\lambda,\beta_0^{\mathrm{c}}}$ }
    \label{fig:Iterative_Bogoliubov_theory}
\end{figure}

We can repeat these steps to move from $\ket{\beta_0',\lambda'}$ to $\ket{\beta_0'',\lambda'}$, $\ket{\beta_0'',\lambda''}$ and so on. For the Bose-Hubbard model, this algorithm converges to the best Gaussian state
\begin{align}
    \ket{\psi_{\mathrm{g}}}=\ket{\lambda^{\mathrm{g}},\beta_0^{\mathrm{g}}}=\lim_{n\to\infty}\ket{\beta_0^{(n)},\lambda^{(n)}}\,,
\end{align}
\ie iterated Bogoliubov theory gives the same approximate ground state as imaginary time evolution on the variational class of all Gaussian states. In fact, we can run into similar troubles as for imaginary time evolution: We could get stuck in a local minima, \ie a state that is the ground state of its mean field Hamiltonian without being the global minimum among all Gaussian states. Another problem of iterated Bogoliubov theory could arise if we encounter $V^{(n)}_k>2U^{(n)}_k$ at some step $n$, in which case the mean field Hamiltonian $[\hat{H}]_{\ket{\psi^{(n)}}}$ would not be bounded from below and consequently we could not compute the next ground state.

\begin{figure}[t]
\centering
  \begin{tikzpicture}
  \draw (0,0) node[inner sep=0pt]{\includegraphics[width=\linewidth]{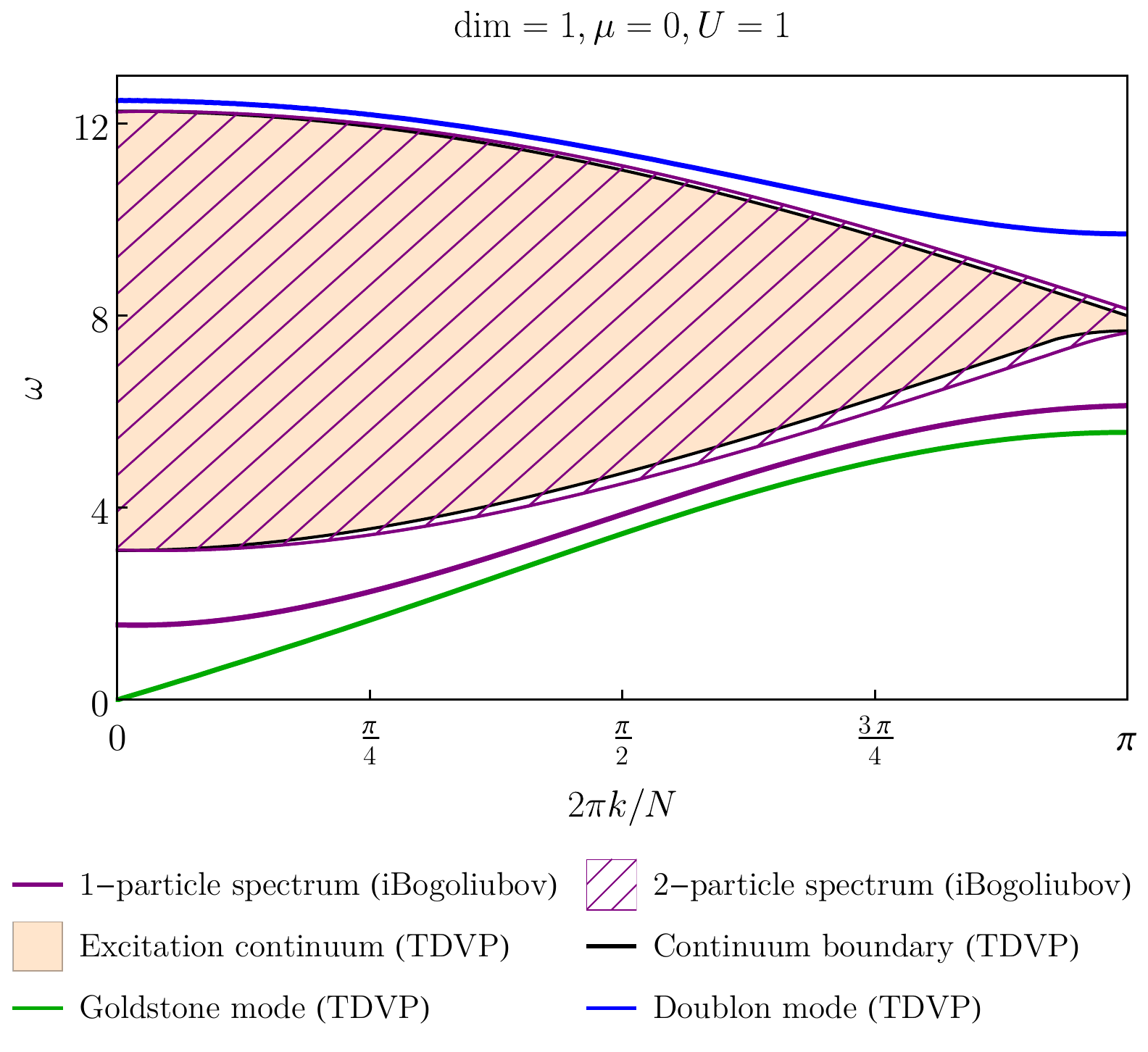}};
  \end{tikzpicture}
  \ccaption{Comparison of Gaussian TDVP (labelled as TDVP) with iterated Bogoliubov theory}{We compare the data of Gaussian TDVP from figure~\ref{fig:excitation_spectrum} with the 1- and 2-particle spectrum of the non-interacting Hamiltonian of $[\hat{H}]_{\ket{\psi_{\mathrm{g}}}}$ from~\eqref{eq:HiBogoliubov}, which we obtained from the iterated Bogoliubov theory (labelled as iBogoliubov).}
  \label{fig:IteratedBogoliubovSpectrum}
\end{figure}

At $\ket{\psi_{\mathrm{g}}}$, we find the quadratic Hamiltonian
\begin{align}
    [\hat{H}]_{\ket{\psi_{\mathrm{g}}}}=E_{\ket{\psi_0}}+\sum_k \mathcal{E}^{\mathrm{g}}_k\,(\delta\hat{B}^{\mathrm{g}}_k)^{\dagger}\delta\hat{B}^{\mathrm{g}}_k\label{eq:HiBogoliubov}
\end{align}
with $\delta \hat{B}_k^{\mathrm{g}}=\lim_{n\to\infty}\delta\hat{B}_k^{(n)}$, that provides a self-consistent generalization of traditional Bogoliubov theory. In particular, we find $\ket{\psi_{\mathrm{g}}}$ is the ground state of its own quadratic Hamiltonian $[\hat{H}]_{\ket{\psi_{\mathrm{g}}}}$. Moreover, the ground state energy of $[\hat{H}]_{\ket{\psi_{\mathrm{g}}}}$ coincide with the expectation value $\braket{\psi_{\mathrm{g}}|\hat{H}|\psi_{\mathrm{g}}}$ or, put differently, we find $\lim_{n\to\infty}\Delta^{(n)}=0$. However, when  looking at the spectrum $\mathcal{E}_k^{\mathrm{g}}=\lim_{n\to\infty}\mathcal{E}_k^{(n)}$ around the Gaussian ground state approximation $\ket{\psi_{\mathrm{g}}}$, we find a gap $\mathcal{E}_0^{\mathrm{g}}>0$, \ie the 1-particle spectrum computed from iterated Bogoliubov theory does not capture the massless Goldstone mode and is therefore worse than traditional Bogoliubov theory. However, we can use the quadratic Hamiltonian $[\hat{H}]_{\ket{\psi_{\mathrm{g}}}}$ to compute a 2-particle spectrum based on the assumption that the particles with dispersion relation $\mathcal{E}_k^{\mathrm{g}}$ do not interact. When comparing the resulting 2-particle continuum with the one from Gaussian TDVP, we find good agreement--in particular in the region around $k=0$.

\section{Computation of Gaussian ground state approximation}\label{app:approximate_ground_state}
We review the underlying analytical and semi-analytical methods associated to Section~\ref{sec:Gaussian_ground_state_approximation}, that enabled us to compute the best Gaussian state, \ie the Gaussian state $\ket{\psi_{\mathrm{g}}}$ with the lowest energy expectation value with respect to $\hat{H}$.
We can restrict ourselves to searching for the ground state in the translationally invariant submanifold, which is parametrized by $\beta_0$ and $\lambda_k:=\lambda_{0,k}$. Due to the $\mathrm{U}(1)$ invariance, it is always possible to find a ground state in which both these parameters are real. It turns out to be very convenient to parametrize the state in terms of the Bogoliubov parameters $\beta_k$, $u_k=\cosh \lambda_k$ and $v_k=\sinh{\lambda_{k}}$ such that
\begin{align}
\begin{split}
    \delta \hat{B}_k&=\mathcal{U}(\beta,\lambda)\,\hat{b}_k\,\mathcal{U}^\dag(\beta,\lambda)\\
    &=u_k (\hat{b}_k-\beta_k) - v_k (\hat{b}^\dag_{-k}-\beta_k)
\end{split}
\end{align}
will annihilate the Gaussian state $\ket{\beta,\lambda}=\mathcal{U}(\beta,\lambda)\ket{0}$.

The stationary point $\ket{\psi_{\mathrm{g}}}$ of the imaginary time evolution is characterized by vanishing $\mathbb{P}_{\ket{\psi_{\mathrm{g}}}}(-\hat{H})\ket{\psi_{\mathrm{g}}}$, which translates into the conditions
\begin{align}
    \braket{0|\hat{b}_0 \mathcal{U}^\dag(\beta,\lambda) H |\psi(\beta,\lambda)}&=0\,, \\
    \braket{0|\hat{b}_k \hat{b}_{-k}\mathcal{U}^\dag(\beta,\lambda) H |\psi(\beta,\lambda)}&=0\,.
\end{align}
Rewriting these conditions in terms of our parameters $(\beta_0,\lambda_k)$ gives
\begin{align}
    \varepsilon_0+\frac{U}{N}(\beta_0^2+A+2B)&=0\,,\\
    \left[\varepsilon_k+\frac{2U}{N}(\beta_0^2+B)\right]u_k v_k + \frac{U}{2N}(\beta_0^2+A) (u_k^2+v_k^2)&=0\,,
    \label{eq:ground_state_equations}
\end{align}
where we defined $A=\sum_k u_k v_k$ and $B=\sum_k v_k^2$ and $\varepsilon_k$ is the dispersion relation in~\eqref{eq:nonint_epsilon}. Equations~\eqref{eq:ground_state_equations} are solved by
\begin{align}
    {\beta^{\mathrm{g}}_0}^2&=-\frac{N\varepsilon_0}{U}-A-2B\,,\label{eq:groundstate_solutions_b}\\
    u^{\mathrm{g}}_k&=\frac{1}{\sqrt{2}} \sqrt{{(1-T_k^2)}^{-\frac{1}{2}}+1}\,,\label{eq:groundstate_solutions_u}\\
    v^{\mathrm{g}}_k&=\frac{1}{\sqrt{2}}\,\mbox{sign} \, T_k \, \sqrt{{(1-T_k^2)}^{-\frac{1}{2}}-1}\,,
    \label{eq:groundstate_solutions_v}
\end{align}
where we introduced the convenient parameter
\begin{align}
    T_k=-\left(1+\frac{2U}{N}\frac{B}{\varepsilon_0}\right) {\left(2-\frac{\varepsilon_k}{\varepsilon_0}+\frac{2U}{N} \frac{A+B}{\varepsilon_0}\right)}^{-1}\,.
\end{align}
This expression for the solution depends on the final values of the quantities $A$ and $B$ which have to be obtained from the coupled equations
\begin{widetext}
\begin{align}
    A&=\frac{1}{2}\sum_k\frac{2B U+N\varepsilon_0}{\sqrt{\left(2U(A +2B)+N(3\varepsilon_0-\varepsilon_k)\right)\left(2AU +N (\varepsilon_0-\varepsilon_k)\right)}}\,,\\
    B&=\frac{1}{2}\sum_k\frac{N(\varepsilon_k-2\varepsilon_0)-2(A+B)U}{\sqrt{\left(2U(A+2B)+N(3\varepsilon_0-\varepsilon_k)\right)\left(2AU+N(\varepsilon_0-\varepsilon_k)\right)}}-\frac{N}{2}\,,\label{eq:AandB}
\end{align}
\end{widetext}
which can be solved numerically efficiently independently of the dimensionality of the system and with a linear dependence on the system size.

We can similarly express the energy $E$ and particle density $n$ of a Gaussian state $\ket{\psi}$ in terms of $\beta_0$, $u_k$, $v_k$, $A=\sum_{k}u_kv_k$ and $B=\sum_kv_k^2$ as
\begin{align}
\begin{split}
    E&=\sum_{k}\varepsilon_k v_k^2-\frac{N\varepsilon_0^2}{2U}-(A+2B)\varepsilon_0-\frac{U}{N}(2A+B)B\,,\\
    n&=\frac{\braket{\hat{N}}}{N}=-\frac{\varepsilon_0}{U} -\frac{A+B}{N}\,.
\end{split}\label{eq:ground_state_properties}
\end{align}
In particular, we can use $\beta_0^{\mathrm{g}}$, $u_k^{\mathrm{g}}$ and $v_k^{\mathrm{g}}$ to compute $E_{\ket{\psi_{\mathrm{g}}}}$.

\section{Linearized equations of motion}\label{app:linearized_eom}
For calculating the matrix $K$ it is convenient to consider the tangent plane as a real vector space of twice the dimensions compared to the complex one, \ie we have
\begin{align}
\begin{split}
    \mathcal{T}_{\ket{\psi}}\mathcal{M}&=\left\{\mathcal{U}(x)\hat{b}^\dag_k\ket{0},\mathcal{U}(x)\hat{b}^\dag_{k-q}\hat{b}^\dag_{q}\ket{0},\right.\\
    &\hspace{50pt}\left.\ii\mathcal{U}(x)\hat{b}^\dag_k\ket{0},\ii\mathcal{U}(x)\hat{b}^\dag_{k-q}\hat{b}^\dag_{q}\ket{0}\right\}\label{eq:real_tangent_frame}
\end{split}
\end{align}
with \emph{real} linear combinations of $N(N+3)$ tangent vectors. The generator of linearized time evolution is then
\begin{align}
    K=-\ii T^{-1} S T
\end{align}
where the matrix $T$ defined in~\eqref{eq:matrix T}, taking the subdivision into blocks to refer to the split between real and imaginary vectors of~\eqref{eq:real_tangent_frame}.

The matrix $S$ is block diagonal with each block $S_k$ referring to a fixed total momentum. Each block can be written as the sum of a diagonal matrix and a rank 5 matrix, that is $S_k=E+CR$, with
\begin{align}
    &E=\begin{pmatrix}
        E_k&0&0&0 \\
        0&\Delta_{q,\tilde{q}}&0&0\\
        0&0&-E_k&0\\
        0&0&0&-\Delta_{q,\tilde{q}}
    \end{pmatrix},\, R=\begin{pmatrix}
        1 & 0 & 0 & 0\\
        0 & 0 & 1 & 0 \\
        0  & a_{k,\tilde{q}} & 0 & a_{k,\tilde{q}} \\
        0  & b_{k,\tilde{q}} & 0 & c_{k,\tilde{q}} \\
        0  & c_{k,\tilde{q}} & 0 & b_{k,\tilde{q}}
    \end{pmatrix},\nonumber\\
    &C=\begin{pmatrix}[1.2]
        0&G_k&\frac{2U}{N}\beta_0(u_k+v_k)&\frac{U}{N} \beta_0v_k&\frac{U}{N}\beta_0u_k\\
         F_k & \bar{F}_k & \frac{2U}{N} a_{k,q} & \frac{U}{2N} b_{k,q} & \frac{U}{2N} c_{k,q} \\
        -G_k & 0 & -\frac{2U}{N}\beta_0 (u_k+v_k) & -\frac{U}{N}\beta_0 u_k & -\frac{U}{N}\beta_0 v_k \\
        -\bar{F}_k & -F_k &  -\frac{2U}{N} a_{k,q} & -\frac{U}{2N} c_{k,q} & -\frac{U}{2N} b_{k,q}
    \end{pmatrix}\nonumber
\end{align}
with newly introduced parameters
\begin{align}
E_k&=(\varepsilon_k+\frac{2U}{N}(\beta_0^2+B))(u_k^2+v_k^2)+\frac{2U}{N}(\beta_0^2+A) u_k v_k, \nonumber \\
\Delta_{q,\tilde{q}}&= (\delta_{q,\tilde{q}}+\delta_{q,-\tilde{q}})(E_{\frac{k}{2}+\tilde{q}}+E_{\frac{k}{2}-\tilde{q}}), \nonumber\\
G_k&=2 (\varepsilon_k+\frac{2U}{N}(\beta_0^2+B)) u_k v_k +\frac{U}{N}(\beta_0^2+A)(u_k^2+v_k^2), \nonumber\\
F_k&=\frac{U}{N} \beta_0 \left[2 a_{k,q}(u_k+v_k)+b_{k,q} v_k +c_{k,q} u_k\right], \nonumber\\
\bar{F}_k &=\frac{U}{N} \beta_0 \left[2 a_{k,q}(u_k+v_k)+b_{k,q} u_k +c_{k,q} v_k\right], \nonumber\\
a_{k,q}&=u_{\frac{k}{2}+q} v_{\frac{k}{2}-q}+u_{\frac{k}{2}-q} v_{\frac{k}{2}+\tilde{q}}, \nonumber \\
b_{k,q}&=2 v_{\frac{k}{2}+q} v_{\frac{k}{2}-q}, \nonumber \\
c_{k,q}&=2 u_{\frac{k}{2}+q} u_{\frac{k}{2}-q},\nonumber%
\end{align}
where $u_k$ and $v_k$ have to be evaluated at the solutions corresponding to the ground state approximation defined in Appendix~\ref{app:approximate_ground_state}, \ie  at $\beta_0^{\mathrm{g}}$, $u^{\mathrm{g}}_k$ and $v^{\mathrm{g}}_k$ from~(\ref{eq:groundstate_solutions_b}-\ref{eq:groundstate_solutions_v}), and $\varepsilon_k$ is the dispersion relation in~\eqref{eq:nonint_epsilon}.

Given the simple structure of the blocks $S_k$, it is easy to diagonalize them numerically. Their eigenvalues are the zeros of the function $f(\omega):=\det[1+R{(E-\omega)}^{-1}C]$. Evaluation only scales linearly with the system size $N$ and moreover, we can characterize analytically some properties of the spectrum in the thermodynamic limit.

More specifically, the function $f(\omega)$ presents a series of poles, given by the diagonal elements of $E$. Its zeros (\ie the eigenvalues of the system) are positioned one in between each pair of subsequent poles. One subset of the poles, that is the diagonal elements of $\Delta$, for $N\to\infty$ come closer together, creating in the thermodynamic limit a continuous line. The zeros that are in between such poles will therefore also come together to a continuum that represents the continuum in the spectrum of $S_k$. The boundaries of this continuum can thus be inferred by computing the values of the minimal and maximal diagonal elements of $\Delta$. In particular, we can identify the minimum, given by $2 E_k$, with the Higgs excitation mode. In order to give an expression for the Higgs gap at zero momentum, we need to evaluate $2 E_0$.

We want to do this at constant filling $n$, which is equivalent to imposing $\varepsilon_0=-U\left(n+\frac{A+B}{N}\right)$, due to~\eqref{eq:ground_state_properties}. Substituting this condition into equation~\eqref{eq:AandB}, we find equations for $A$ and $B$ at fixed $n$. These equations admit constant solutions in the limit $U\to 0$. Inserting these solutions in the expression for $E_0$, we find the asymptotics of the Higgs gap at constant density for $U\to 0$
\begin{align}
    2E_0\sim  \alpha(N,n)\,U\quad\text{as}\quad U\to 0\,. 
\end{align}
The function $\alpha$ has a complicated analytical expression that admits the large $N$ asymptotics
\begin{align}
    \alpha(N,n)\sim 2\sqrt[3]{2} n^{\frac{2}{3}} N^{-\frac{1}{3}}\quad\text{as}\quad N\to\infty\,.
\end{align}

\section{Linear response theory}\label{app:linear_response_theory}
We are interested in computing the linear variation $\delta V(t) = \frac{d}{d\lambda} \bra{\psi_\lambda(t)}\hat{V}\ket{\psi_\lambda(t)}_{\lambda=0}$ due to the perturbation $\lambda\varphi(t)\hat{V}$ of the Hamiltonian. With respect to a set of real coordinates $x^a$ of our variational manifold and the corresponding basis $\ket{V_a}=\frac{\partial}{\partial x^a}\ket{\psi(x^{\mathrm{g}})}$ (where $x^{\mathrm{g}}$ are the coordinates of the stationary point $\ket{\psi_{\mathrm{g}}}$), \eg the ones of~\eqref{eq:real_coordinates} and~\eqref{eq:real_tangent_frame}, we find
\begin{align}
    \delta V(t) = dV_a \, \delta \psi^a(t)\,,
    \label{eq:variation_of_response}
\end{align}
with $dV_a=\frac{\partial}{\partial x^a} \bra{\psi(x)}\hat{V}\ket{\psi(x)}$ and 
\begin{align}
    \delta\psi^a(t)\ket{V_a}=\ket{\delta\psi(t)}=\frac{d}{d\lambda}\Bigg|_{\lambda=0}\hspace{-5mm}\ket{\psi_\lambda(t)}\,.
\end{align}
Put differently, $\delta\psi^a(t)$ is the component of the tangent vector $\ket{\delta\psi(t)}$ in the direction $\ket{V_a}$. Note that in~\eqref{eq:variation_of_response} as well as in the rest of this appendix we use the Einstein convention for indices, whereby it is understood that all repeated indices are summed over.

Furthermore, we can introduce the components $v^a$ of the tangent vector
\begin{align}
    v^a\ket{V_a}=\mathbb{P}_{\ket{\psi_{\mathrm{g}}}}(-\ii\hat{V})\ket{\psi_{\mathrm{g}}}\,
\end{align}
representing the linear perturbation of $\ket{\psi_{\mathrm{g}}}$ due to $\hat{V}$.

The key result of the following paragraph is
\begin{align}
    \delta\psi^a(t)=\int_{-\infty}^t dt'\, \varphi(t')\,(d\Phi_{t-t'})^a{}_b \, v^b\,,\label{eq:linear_response}
\end{align}
where $d\Phi_t$ is the linearized flow around the stationary point $\ket{\psi_{\mathrm{g}}}$, often also referred to as push-forward map.
The linear response $\delta\psi^a(t)$ at time $t$ can be understood as functional $\delta\psi^a(t)[\varphi]$ of $\varphi: [-\infty,t]\to\mathbb{R}$ that modulates the perturbation $\hat{V}$ for previous times. We can therefore write
\begin{align}
    \delta\psi^a(t)[\varphi]=\int_{-\infty}^t dt'\, \varphi(t')\frac{\delta\psi^a(t)}{\delta \varphi(t')}\,,\label{eq:variation_integral}
\end{align}
which states that the $\delta\psi^a(t)$ is the superposition of all linear responses
\begin{align}
    \frac{\delta\psi^a(t)}{\delta \varphi(t')}=\frac{d}{d\lambda}\Bigg|_{\lambda=0}\hspace{-5mm}\ket{\psi_\lambda(t)}_{\varphi(t)=\delta(t-t')}\label{eq:variation_response}
\end{align}
due to a perturbation with $\varphi(t)=\delta(t-t')$. This scenario can be evaluated explicitly to be given by
\begin{align}
    \ket{\psi_\lambda(t)}_{\varphi(t)=\delta(t-t')}=\Phi_{t-t'}\Phi^{\hat{V}}_\lambda\ket{\psi_{\mathrm{g}}}\,.\label{eq:kick}
\end{align}
Here, the state $\ket{\psi_{\mathrm{g}}}$ is unaffected until time $t'$. At this point, it instantaneously kicked to the new state $\ket{\psi_\lambda(t')}=\Phi^{\hat{V}}_\lambda\ket{\psi_{\mathrm{g}}}=\lambda v^a\ket{V_a}$, where $\Phi^{\hat{V}}_\lambda$ represents the projected time evolution with respect to the Hamiltonian $\lambda \delta(t-t') \hat{V}$. After the kick, time evolution continuous to be unperturbed and therefore given by $\Phi_{t-t'}$ until the time $t$ that we are interested in. Evaluating the derivative with respect to $\lambda$ in~\eqref{eq:variation_response} therefore gives
\begin{align}
    \frac{\delta\psi^a(t)}{\delta \varphi(t')}=(d\Phi_{t-t'})^a{}_bv^b\,,\label{eq:variation_explicit}
\end{align}
where $d\Phi_{t-t'}$ is the linearization of $\Phi_{t-t'}$ and $v^b\ket{V_b}=\frac{d}{d\lambda}|_{\lambda=0}\Phi^{\hat{V}}_\lambda\ket{\psi_{\mathrm{g}}}$. Plugging~\eqref{eq:variation_explicit} into~\eqref{eq:variation_integral} gives~\eqref{eq:linear_response}.

By observing that
\begin{align}
     v^a={\Omega}^{ab} dV_b(\psi),
 \end{align}
we finally have the expression
\begin{align}
     \delta V(t)=\int_{-\infty}^t dt'\, \left[ dV_a \, \left(d\Phi_{t-t'}\right)^a{}_b \, \Omega^{bc} \, dV_c \right] \varphi(t')\,,
     \label{eq:response_time_domain}
\end{align}
where we represent the symplectic form as
\begin{align}
    \Omega=\left(\begin{array}{cc} 0 & \id \\ -\id & 0 \end{array} \right)\,. \label{eq:Omega}
\end{align}

Next, we consider the Fourier transformed response function, that can be written as 
\begin{align}
	\delta V(\omega)&\equiv\int dt \,  e^{-i\omega t}\delta V(t)  \nonumber\\
	&=\tilde{\varphi}(\omega) \; \int d\omega'\, \frac{Z_V(\omega')}{\omega'-\omega + i 0^{+}}.
	\label{eq:response_frequency_domain}
\end{align}
We define $Z_V(\omega)$ as the spectral response function.

In order to obtain the expression~\eqref{eq:response_frequency_domain}, it is useful to give a spectral decomposition of the linear operator $d\Phi_t$. As mentioned, if we consider as the initial state of our evolution the approximate ground state $\ket{\psi_{\mathrm{g}}}$, $d\Phi_t$ becomes a linear map from the tangent plane $\mathcal{T}_{\ket{\psi_{\mathrm{g}}}}\mathcal{M}$ onto itself, given by the exponential of the generator $K$, \ie we have
\begin{align}
    d\Phi_t=e^{Kt}\,.
\end{align}
The generator $K$ was computed in Appendix~\ref{app:linearized_eom} and, as discussed before, its eigenvectors $e^a(\omega)$ appear in complex conjugate pairs satisfying
\begin{align}
    K^a{}_be^{b}(\omega)&=+\ii\omega e^a(\omega)\,, \nonumber\\
    K^a{}_be^{*b}(\omega)&=-\ii\omega e^{*a}(\omega)\,.
\end{align}
This means that $K$ can be decomposed as
\begin{align}
    K=O \, \bigoplus_i \left(\begin{array}{cc} 0 & -\omega_i \\ \omega_i & 0 \end{array} \right) O^{-1},
\end{align}
with
\begin{align}
	O=\left(\begin{array}{c|c|c|c|c} \mbox{Re}\,e(\omega_1) & -\mbox{Im}\,e(\omega_1) & \cdots & \mbox{Re}\,e(\omega_n) & -\mbox{Im}\,e(\omega_n) \end{array} \right),
\end{align}
where the $\omega_i$ are all taken to be positive. 

Using this, the object in the square brackets in~\eqref{eq:response_time_domain} can be written as
\begin{align}
\begin{split}
    &\sum_i\left(d{V}_+ (\omega_i), d{V}_- (\omega_i) \right) \, \left(\begin{array}{cc} \cos\omega_i t & -\sin\omega_i t \\ \sin\omega_i t & \cos\omega_i t \end{array} \right)\\
	&\qquad \times\left(\begin{array}{cc} 0 & -\delta(\omega_i) \\ \delta(\omega_i) & 0 \end{array} \right) \left(\begin{array}{c} d{V}_+ (\omega_i) \\ d{V}_- (\omega_i) \end{array} \right)
\end{split}\\
\begin{split}
    &=\ii \int_{0}^\infty d\omega \, \frac{d{V}_- (\omega)^2+d{V}_+ (\omega)^2}{2} \delta(\omega) e^{- \ii\omega t}\\
    &\qquad -\ii \int_{0}^\infty d\omega \,  \frac{d{V}_- (\omega)^2+d{V}_+ (\omega)^2}{2} \delta(\omega) e^{\ii\omega t}\,,
\end{split}
\end{align}
where we introduced the terms
\begin{align}
    d{V}_+ (\omega)&=\mathrm{Re}\,e^a(\omega) \, dV_a\,,\\
    d{V}_- (\omega)&=-\mathrm{Im}\,e^a(\omega) \, dV_a\\
    \delta(\omega)&=[\mathrm{Im}\,e^a(\omega)] \Omega_{ab} [\mathrm{Re}\,e^b(\omega)]\,,\label{eq:delta}
\end{align}
such that the vectors $e(\omega)$ are normalized to satisfy $\delta(\omega)=\pm1$.

Taking the Fourier transform as in~\eqref{eq:response_frequency_domain}, we find
\begin{align}
    Z_V(\omega)&= \mbox{sign}(\omega) \,  \frac{d{V}_- (|\omega|)^2+d{V}_+ (|\omega|)^2}{2} \delta(|\omega|)\nonumber\\
    &=\mbox{sign}(\omega) \,  \frac{{\left|e^a(\omega) \, dV_a \right|}^2}{2} \delta(|\omega|).
\end{align}

The generator $K^a{}_b$ is a linear map on the tangent space, taken as real vector space. It is not complex-linear, \ie $K^a{}_b$ does not commute with $J^a{}_b$, which is the linear map representing multiplication with the imaginary unit
\begin{align}
    J^a{}_bv^b\ket{V_a}=\ii v^a\ket{V_a}\,.
\end{align}
In particular, this implies that $K$ is not anti-Hermitian with respect to the standard inner product, which means that $d\Phi_t=e^{tK}$ is not unitary.

However, it is easy to find an alternative inner product on the tangent space that turns $K$ into an anti-Hermitian operator and is therefore preserved by the linearized flow $d\Phi_t$. Given two vectors $\ket{x}=x^a\ket{V_a}$ and $\ket{y}=y^a\ket{V_a}$, we can define the inner product
\begin{align}
    {\braket{x|y}}_{\tilde{g}}:=x^{*a} \tilde{g}_{ab} y^{b}\quad\text{with}\quad \tilde{g}_{ab}=\Omega^{-1}_{ac} K^{\mathrm{c}}{}_b\,,
    \label{eq:inner_product_M}
\end{align}
which is a well-defined Hermitian inner product. With respect to this inner product, $K$ is anti-Hermitian, \ie $K^\dagger=-K$, which implies that $d\Phi_t=e^{tK}$ is unitary.

The modified inner product~\eqref{eq:inner_product_M} on the tangent plane can be used to calculate overlaps between vectors, consistent with the unitary time evolution. In particular, the ovelaps between the evolving perturbation and the 1-particle subspace of the tangent plane shown in Figure~\ref{fig:time_evolution} are calculated with this inner product.

\section{Random phase approximation}\label{app:feynman_diagrams}
\def\propag[#1,#2,#3,#4](#5,#6){\draw[#1,#2] (#5,#6) -- (#5+1,#6) node[anchor=south, pos=0.5]{#3} #4;}
\def\pibub[#1,#2,#3,#4](#5,#6){		
	\draw[db,#1](#5,#6) arc (160:46:0.5) node[anchor=west, pos=0]{#2};
	\draw[db,#1] (#5,#6) arc (200:314:0.5);
	\draw[db,->-] (#5+1.25,#6)--(#5+2.3,#6) #4;
	\filldraw[color=black, fill=black!50] (#5+1,#6) circle (0.3) node[anchor=south, inner sep=14]{#3};
}
\def\pizero[#1](#2,#3){			
	\draw[#1](#2,#3) arc (160:20:0.5) node[anchor=south, pos=0.5]{$\Pi_0$};
	\draw[#1] (#2,#3) arc (200:340:0.5)
}
\def\piu{node[anchor=west, pos=1, inner sep=10]{$+$}}
\def\uguale{node[anchor=west, pos=1, inner sep=10]{$=$}}
\begin{figure*}[t]
\centering
\begin{tikzpicture}
    \propag[db,->-,$G$,\uguale](0,0)
	\propag[,->-,$G_0$,\piu](2,0)
	\propag[ ,->-,$G_0$, ](4,0)
	\pibub[->-,$g$,$\Pi$,\piu](5,0)	
	\propag[ ,->-,$G_0$, ](8.3,0)
	\pibub[-<-,$\bar{g}$,$\bar{\Pi}$, ](9.3,0)

\begin{scope}[yshift = .7cm]
	\propag[db,<->,$F$,\uguale](0,-2)
	\propag[ ,-<-,$G_0$, ](2,-2)
	\pibub[->-,$\bar{g}^*$,$\Pi$,\piu](3,-2)	
	\propag[ ,-<-,$G_0$, ](6.3,-2)
	\pibub[-<-,$g^*$,$\bar{\Pi}$, ](7.3,-2)
\end{scope}

\begin{scope}[yshift = 1.4cm]
	\pibub[->-, ,$\Pi$,\uguale](0-1.3,-4)
	\pizero[->-](3.3-1.3,-4);
	\pibub[->-,$\Theta$,$\Pi$,\piu](4.25-1.3,-4)
	\pizero[->-](4.25+3.3-1.3,-4);
	\pibub[-<-,$\bar{\Theta}$,$\bar{\Pi}$,\piu](4.25+3.3+0.95-1.3,-4)
	\pizero[->-](8.5+3.3-1.3,-4) node[anchor=east, pos=1]{$g^*$};
	\propag[db,->-,$G$,\piu](8.5+3.3+0.95-1.3,-4)
	\pizero[->-](12.75+1.95-1.3,-4) node[anchor=east, pos=1]{$\bar{g}$};
	\propag[db,<->,$F$, ](12.75+1.95+0.95-1.3,-4)
\end{scope}

\begin{scope}[yshift = 2.1cm]
	\pibub[-<-, ,$\bar{\Pi}$,\uguale](0-1.3,-6)
	\pizero[-<-](3.3-1.3,-6);
	\pibub[->-,$\bar{\Theta}^*$,$\Pi$,\piu](4.25-1.3,-6)
	\pizero[-<-](4.25+3.3-1.3,-6);
	\pibub[-<-,$\Theta^*$,$\bar{\Pi}$,\piu](4.25+3.3+0.95-1.3,-6)
	\pizero[-<-](8.5+3.3-1.3,-6) node[anchor=east, pos=1]{$\bar{g}^*$};
	\propag[db,->-,$G$,\piu](8.5+3.3+0.95-1.3,-6)
	\pizero[-<-](12.75+1.95-1.3,-6) node[anchor=east, pos=1]{$g$};
	\propag[db,<->,$F$, ](12.75+1.95+0.95-1.3,-6)
\end{scope}
\end{tikzpicture}
\ccaption{Feynman diagrams}{Pictorial representation in terms of Feynman diagrams of the terms included in equations~(\ref{eq:corrG}-\ref{eq:corrPb}) for the correlation functions}
\label{fig:feyndiag}
\end{figure*}

In this section, we show the fluctuation spectrum from the Gaussian state approach can also be obtained from the standard random phase approximation (RPA) based on the ladder diagram, similarly to what has been done for fermions in~\cite{demler_collective_excitations_1996}.

Around any Gaussian variational state $\ket{\psi}$, characterized by the real numbers $\beta _{0}$, $u_k$ and $v_k$ as described in Appendix~\ref{app:approximate_ground_state}, the Hamiltonian $H=E+\sum_{j=1}^{4}h_{j}$ is decomposed into five normal ordered terms (with respect to the given Gaussian state $\ket{\psi}$) via the Wick theorem. We will here be interested in taking this reference state to be our self-consistent Gaussian ground state approximation $\ket{\psi_{\mathrm{g}}}$ parametrized by real $\beta_0^{\mathrm{g}}$, $u_k^{\mathrm{g}}$ and $v^{\mathrm{g}}_k$ from~(\ref{eq:groundstate_solutions_b}-\ref{eq:groundstate_solutions_v}).

For this particular state, the energy $E$ is given by~\eqref{eq:ground_state_properties}. The linear term $h_{1}$ vanishes. The quadratic term, when written in terms of the Bogoliubov-transformed operators
\begin{align}
    \delta\hat{B}_k=u_k(\hat{b}_k-\beta_k)-v_k(\hat{b}_k-\beta_k)\,,
\end{align}
takes the diagonal form
\begin{align}
    h_2=\sum_k\mathcal{E}_k\delta\hat{B}_k^\dag\delta\hat{B}_k\,,
\end{align}
where the dispersion relation $\mathcal{E}_{k}=\sqrt{U_{k}^{2}-V_k^{2}}$ with 
\begin{align}
U_{k} =\varepsilon _{k}+\frac{2U}{N}(\left\vert \beta
_{0}\right\vert ^{2}+\sum_{q}v_q^2
)\,,\,\,\,
V_k=\frac{U}{N}(\beta_{0}^{2}+\sum_{q}u_q v_q)\,,
\end{align}%
coincides with~\eqref{eq:E'_k}.
\begin{widetext}
\noindent The cubic and quartic terms become
\begin{align}
\begin{split}
h_3&=\frac{U}{N}\beta_{0}\sum_{kp}[(u_{k}u_{p}u_{k+p}+v_{k}v_{p}v_{k+p}+2u_{k}v_{p}u_{k+p}+2u_{k}v_{p}v_{k+p})\delta \hat{B}_{k}^{\dagger }\delta \hat{B} _{p}^{\dagger }\delta \hat{B}_{k+p}\\
&\quad+(u_{k}u_{p}v_{k+p}+v_{k}v_{p}u_{k+p})\delta \hat{B}_{k}^{\dagger }\delta
B_{p}^{\dagger }\delta \hat{B}_{-k-p}^{\dagger }]+\mathrm{H.c.}
\end{split}\\
\begin{split}
h_{4} &=\frac{U}{N}\sum_{k,p,q}[(u_{k+q}u_{p-q}u_{p}u_{k}+v_{k+q}v_{p-q}v_{p}v_{k}+4u_{k+q}v_{p-q}v_{p}u_{k})\delta \hat{B} _{k+q}^{\dagger }\delta \hat{B}_{p-q}^{\dagger }\delta \hat{B}_{p}\delta \hat{B}_{k}\\
&\quad+u_{k+q}u_{p-q}v_{p}v_{k}\delta \hat{B}_{k+q}^{\dagger }\delta \hat{B}_{p-q}^{\dagger }\delta \hat{B} _{-p}^{\dagger }\delta \hat{B}_{-k}^{\dagger }+2(u_{k+q}u_{p-q}v_{k}u_{p}+u_{k+q}v_{p-q}v_{k}v_{p})\delta \hat{B}
_{k+q}^{\dagger }\delta \hat{B}_{p-q}^{\dagger }\delta \hat{B}_{-k}^{\dagger
}\delta \hat{B}_{p}+\mathrm{H.c.}]\,.
\end{split}
\end{align}
We calculate the time-ordered correlation functions%
\begin{align}
G(k,\omega ) &=-\ii\int dte^{\ii\omega t}\left\langle \mathcal{T}\delta \hat{B}_{k}(t)\delta \hat{B} _{k}^{\dagger }\right\rangle\,,\\
F(k,\omega ) &=-\ii\int dte^{\ii\omega t}\left\langle \mathcal{T}\delta \hat{B}_{k}(t)\delta \hat{B}_{-k}\right\rangle\,, \\
\Pi (k,p,\omega ) &=-\frac{\ii}{\sqrt{2}}\int dte^{\ii\omega t}\left\langle\mathcal{T}\delta \hat{B}_{k}(t)(-\ii\delta \hat{B}_{k+p}^{\dagger }\delta \hat{B}_{-p}^{\dagger })\right\rangle\,, \\
\bar{\Pi}(k,p,\omega ) &=-\frac{\ii}{\sqrt{2}}\int dte^{\ii\omega t}\left\langle \mathcal{T}\delta \hat{B}_{k}(t)(\ii\delta \hat{B}_{-k-p}\delta
B_{p})\right\rangle\,,
\end{align}
by treating $h_{3}$ and $h_{4}$ as perturbations. To the 1-loop order, as shown in Figure~\ref{fig:feyndiag}, the correlation functions read%
\begin{align}
G(k,\omega ) &=G_{0}(k,\omega )+\ii G_{0}(k,\omega )\sum_{p}g_{kp}\Pi
(k,p,\omega )-\ii G_{0}(k,\omega )\sum_{p}\bar{g}_{kp}\bar{\Pi}(k,p,\omega )\,, \label{eq:corrG}\\
F(k,\omega ) &=-\ii G_{0}(k,-\omega )\sum_{p}g_{kp}\bar{\Pi}(k,p,\omega )+\ii\sum_{p}\bar{g}_{kp}\Pi (k,p,\omega )G_{0}(k,-\omega )\,,  \label{eq:corrF}\\
\begin{split}
\Pi (k,p,\omega ) &=\Pi _{0}(k,p,\omega )\sum_{q}\Theta _{pq}^{k}\Pi(k,q,\omega )-\Pi _{0}(k,p,\omega )\sum_{q}\bar{\Theta}_{pq}^{k}\bar{\Pi}%
(k,q,\omega )\\
&\qquad-\ii g_{kp}\Pi _{0}(k,p,\omega )G(k,\omega )-\ii\bar{g}_{kp}\Pi _{0}(k,p,\omega
)F(k,\omega )\,,  \label{eq:corrP}
\end{split}\\
\begin{split}
\bar{\Pi}(k,p,\omega ) &=\Pi _{0}(k,p,-\omega )\sum_{q}\Theta _{pq}^{k}\bar{\Pi}(k,q,\omega )-\Pi _{0}(k,p,-\omega )\sum_{q}\bar{\Theta}_{pq}^{k}\Pi
(k,q,\omega )\\
&\qquad+\ii g_{kp}\Pi _{0}(k,p,-\omega )F(k,\omega )+\ii\bar{g}_{kp}\Pi
_{0}(k,p,-\omega )G(k,\omega )\,,  \label{eq:corrPb}
\end{split}
\end{align}
where the free propagators for single and two excitations are $G_{0}(k,\omega )=(\omega-\mathcal{E}_{k}+\ii 0^{+})^{-1}$ and $\Pi _{0}(k,p,\omega)=(\omega -\mathcal{E} _{k+p}-\mathcal{E}_{p}+\ii 0^{+})^{-1}$.
\end{widetext}

The coupling between single and two excitations is described by interaction vertices%
\begin{align}
g_{kp}&=\frac{\sqrt{2}U\beta_{0}}{N}(u_{k+p}u_{p}u_{k}+v_{k+p}v_{p}v_{k}+u_{k+p}v_{p}u_{k}\nonumber\\
&\quad+u_{p}v_{k+p}u_{k}+u_{k+p}v_{p}v_{k}+u_{p}v_{k+p}v_{k})\,,\\
\bar{g}_{kp} &=\frac{\sqrt{2}U\beta_{0}}{N}(u_{k+p}u_{p}v_{k}+v_{k+p}v_{p}u_{k}+v_{k+p}u_{k}u_{p}\nonumber\\
&\quad+v_{p}u_{k}u_{k+p}+\! u_{k+p}v_{k}v_{p}+\!u_{p}v_{k}v_{k+p})
\end{align}%
and the interaction vertices of two excitatations are%
\begin{align}
\begin{split}
\Theta _{pq}^{k}&=\frac{U}{N}(u_{p}u_{k+p}u_{k+q}u_{q}+v_{k+q}v_{q}v_{p}v_{k+p}\\
&\qquad+4u_{k+p}v_{p}u_{k+q}v_{q})\,,
\end{split}\\
\begin{split}
\bar{\Theta}_{pq}^{k}&=\frac{U}{N}(u_{k+q}u_{q}v_{p}v_{k+p}+v_{k+q}v_{q}u_{k+p}u_{p}\\
&\qquad+4u_{k+p}v_{p}u_{k+q}v_{q})\,.
\end{split}
\end{align}
In the compact matrix form, we can read~(\ref{eq:corrG}-\ref{eq:corrPb}) as
\begin{align}
(\omega -\mathbf{M})\,\mathbf{G}=\mathbf{V}
\end{align}%
in the basis $\mathbf{G}=(G(k),F(k),\Pi (k,p),\bar{\Pi}(k,p))^{\intercal}$, where $\mathbf{V}=(1,0,0,0)^{\intercal}$, and the matrix
\begin{align}
\mathbf{M}&=
\left(
\begin{array}{cccc}
\mathcal{E}_{k} & 0 & 0 & 0 \\
0 & -\mathcal{E}_{k} & 0 & 0 \\
0 & 0 & (\mathcal{E}_{k+p}+\mathcal{E}_{p})\delta _{pq} & 0 \\
0 & 0 & 0 & -(\mathcal{E}_{k+p}+\mathcal{E}
_{p})\delta _{pq}
\end{array}%
\right)\nonumber\\
&\qquad\quad+\left(
\begin{array}{cccc}
0 & 0 & \ii g_{kp} & -\ii\bar{g}_{kp} \\
0 & 0 & -\ii\bar{g}_{kp} & \ii g_{kp} \\
-\ii g_{kp} & -\ii\bar{g}_{kp} & \Theta
_{pq}^{k} & -\bar{\Theta}_{pq}^{k} \\
-\ii\bar{g}_{kp} & -\ii g_{kp} & \bar{\Theta}_{pq}^{k} & -\Theta _{pq}^{k}%
\end{array}%
\right)\,.
\end{align}
It can be verified that, if we evaluate the parameters $\beta_0$, $u_k$ and $v_k$ to be the ones corresponding to the best Gaussian state $\ket{\psi_{\mathrm{g}}}$, the matrix $\mathbf{M}$ is exactly the linearized time evolution generator matrix $K$ transfromed into the Bogoliubov basis, whose eigenvalues are the poles of the Green function $\mathbf{G}$ thus determine the fluctuation spectrum.
\end{document}